\documentstyle[amsfonts,amssymb,amsmath,graphicx]{article}

\newcommand{\dsp}{\displaystyle}
\newcommand{\un}{\underline}

\newcommand{\ndt}{\noindent}
\newcommand{\qed}{\hfill $\square$}

\newtheorem{theorem}{Theorem}[section]

\newtheorem{corollary}[theorem]{Corollary}
\newtheorem{remark}[theorem]{Remark}
\newtheorem{proposition}[theorem]{Proposition}

\title{\bf Fundamental solutions of the  Dirac operator in  the Friedmann-Lema\^itre-Robertson-Walker spacetime
}

\author{{\bf Karen Yagdjian}
 }

\begin{document}

\date{}
\maketitle
\thispagestyle{empty}
\vspace{-0.3cm}

\begin{center}
{\it Department of Mathematics,
University of Texas RGV,\\
1201 W.~University Drive,  
Edinburg, TX 78539,
USA }
\end{center}
\medskip

\addtocounter{section}{-1}
\renewcommand{\theequation}{\thesection.\arabic{equation}}
\setcounter{equation}{0}
\pagenumbering{arabic}
\setcounter{page}{1}
\thispagestyle{empty}

\hspace{2cm}\begin{abstract}
\begin{small}
The equation of the spin-$\frac{1}{2}$ particles in the Friedmann-Lema\^itre-Robertson-Walker spacetime is investigated.
The retarded and advanced fundamental solutions to the Dirac operator and generalized Dirac operator as well as the fundamental solutions to the Cauchy problem are written in  explicit form via the fundamental solution of the wave equation in the Minkowski spacetime. 
\medskip

\end{small}
\end{abstract}

\setcounter{equation}{0}
\renewcommand{\theequation}{\thesection.\arabic{equation}}

\section{Introduction}

In this article we derive  fundamental solutions of the Dirac operator in the curved spacetime of the Friedmann-Lema\^itre-Robertson-Walker (FLRW) models of cosmology. More precisely, we derive in  explicit form 
the retarded and advanced fundamental solutions to the Dirac operator as well as the fundamental solutions to the Cauchy problem via  the fundamental solution of the wave equation in the Minkowski spacetime.

The metric tensor in the spatially flat FLRW spacetime is 
\[
(g_{\mu \nu })=    \left (
   \begin{array}{ccccc}
 1& 0& 0   & 0 \\
   0& -a^2(t) &  0 & 0 \\ 
 0 & 0 &  -a^2(t)   & 0 \\
 0& 0& 0   &  -a^2(t) \\
   \end{array} \right),\quad \mu ,\nu =0,1,2,3.  
\] 
We will focus on the de~Sitter space with the scale factor   $a(t)=e^{Ht}  $ (see, e.g., \cite{Moller}) that is modeling the expanding or contracting universe if $H>0$ or $H<0$, respectively. The curvature of this space is $-12H^2$. The Dirac equation in the de~Sitter space is (see, e.g., \cite{Barut-D})
\begin{equation}
\label{DE}
  \dsp 
\left(  i {\gamma }^0    \partial_0   +i e^{-Ht}{\gamma }^1  \partial_1+i  e^{-Ht}{\gamma }^2 \partial_2+i e^{-Ht}{\gamma }^3   \partial_3 +i \frac{3}{2}    H {\gamma }^0     -m{\mathbb I}_4 \right)\psi=f \,,
\end{equation}
where 
the contravariant gamma matrices are   (see,  e.g., \cite[p. 61]{B-Sh})
\begin{eqnarray*}
&  &
 \gamma ^0= \left (
   \begin{array}{ccccc}
   {\mathbb I}_2& {\mathbb O}_2   \\
   {\mathbb O}_2& -{\mathbb I}_2   \\ 
   \end{array}
   \right),\quad 
\gamma ^k= \left (
   \begin{array}{ccccc}
  {\mathbb O}_2& \sigma ^k   \\
  -\sigma ^k &  {\mathbb O}_2  \\  
   \end{array}
   \right),\quad k=1,2,3\,.
 \end{eqnarray*}
Here $\sigma ^k $ are Pauli matrices 
\begin{eqnarray*}
&  &
\sigma ^1= \left (
   \begin{array}{ccccc}
  0& 1   \\
  1& 0  \\  
   \end{array}
   \right), \quad
\sigma ^2= \left (
   \begin{array}{ccccc}
  0& -i   \\
  i& 0  \\  
   \end{array}
   \right),\quad
\sigma ^3= \left (
   \begin{array}{ccccc}
  1& 0   \\
  0&-1 \\  
   \end{array}
   \right)\,,
\end{eqnarray*}
and  ${\mathbb I}_n $, ${\mathbb O}_n $ denote the $n\times n$ identity and zero matrices, respectively. In this article we present in explicit forms   the fundamental solutions of the Dirac operator 
in the de~Sitter spacetime. 

The general approach to the study of the Dirac operator 
in the curved spacetime can be described as follows. (See, e.g., \cite[Sec.5.6]{Parker}.)  Denote the Lorenzian metric tensor
\begin{eqnarray*}
 (\eta _{\mu \nu })= \left (
   \begin{array}{ccccc}
   1& 0   \\
   0&   -{\mathbb I}_3  \\ 
   \end{array}
   \right)
\end{eqnarray*}
and by $\un{\gamma }^\mu (x)$ the  matrices, which  
are defined by
\[
\un{\gamma }^\mu (x)\un{\gamma }^\nu  (x)+\un{\gamma }^\nu  (x)\un{\gamma }^\mu (x)=2 g^{\mu \nu }(x)\,, 
\]
while $\un{\gamma }_\nu  (x)= g_{\nu \mu }(x)\un{\gamma }^\mu (x)$. Here and henceforth,  Einstein summation convention over repeated indexes is employed. 
The covariant derivative of a spinor field $\psi $ is
$$
\nabla_{\mu} \psi \equiv\left(\partial_{\mu}-\Gamma_{\mu}\right) \psi\,,
$$
where the spinorial affine connections
$\Gamma _\mu (x)$ are matrices, which are defined by the annihilating  of the covariant derivative of the
$\un{\gamma }$-matrices,
$$
\nabla_{\mu} \underline{\gamma}_{\nu} \equiv \partial_{\mu} \underline{\gamma}_{\nu}-\Gamma^{\lambda}{}_{\mu \nu} \underline{\gamma}_{\lambda}-\Gamma_{\mu} \underline{\gamma}_{\nu}+\underline{\gamma}_{\nu} \Gamma_{\mu}=0\,,
$$
and $ \Gamma^{\lambda}{}_{\mu \nu}$ are affine connections determined by the metric $g $.
Then the covariant Dirac equation in the curved spacetime reads
\[
\left(i \un{\gamma}^{\mu}(x) \nabla_{\mu}-m\right) \psi(x)=0\,.
\]
The following identity
\begin{equation}
\label{0.3}
\underline{\gamma}^{\mu} \nabla_{\mu}\left(\underline{\gamma}^{\nu} \nabla_{\nu} \psi\right)
=\frac{1}{\sqrt{|g|}} \nabla_{\mu}\left(\sqrt{|g|} g^{\mu \nu} \nabla_{\nu}\psi\right)-\frac{1}{4} R\psi \,,
\end{equation}
where $|g|=|\mbox{\rm det}\,(g_{\mu \nu})| $, establishes a relation between the Dirac operator,  spinorial D'Alembert operator and the scalar curvature $R$. (See, e.g., \cite[Eq. (74)]{Schrodinger},
\cite[Sec.~3.9, 5.6]{Parker}.) Hence, if ${\mathcal E}_{KG}(x,x') $ is a fundamental solution to the spinorial Klein-Gordon operator,
\begin{eqnarray}
\label{0.4}
  \left(\frac{1}{\sqrt{|g|}} \nabla_{\mu}\left(\sqrt{|g|} g^{\mu \nu} \nabla_{\nu}\right)-\frac{1}{4} R+m^{2}\right) {\mathcal E}_{KG}\left(x, x^{\prime}\right)=\delta\left(x, x^{\prime}\right) {\mathbb  I}_4\,,
\end{eqnarray}
then
$$
{\mathcal E}\left(x, x^{\prime}\right) :=-\left(i \underline{\gamma}^{\mu} \nabla_{\mu}+m\right) {\mathcal E}_{KG}(x,x')\left(x, x^{\prime}\right)
$$
is the fundamental solution to the Dirac operator:
$$
\left(i \un{\gamma}^{\mu} \nabla_{\mu}-m\right){\mathcal E}\left(x, x^{\prime}\right)=\delta\left(x, x^{\prime}\right) {\mathbb  I}_4\,.
$$ 
In this paper we modify the factorization (\ref{0.3}).  For   the first factor we put the operator of (\ref{DE}) and  then choose   an appropriate  second factor that leads to the diagonal $4\times 4 $ operator matrix containing scalar Klein-Gordon operators with the modified complex-valued mass terms. Thus, in order to construct  fundamental solution to the Dirac operator the only things that remain  are to find an explicit form for the fundamental solution to the Klein-Gordon operator in the curved spacetime and an explicit form for the spin connection.  Obviously,  the Klein-Gordon operator of (\ref{0.4}) has  variable coefficients with the spinorial structure  that makes difficult finding of the explicit representation of the fundamental solutions.  
 The explicit form  of the fundamental solutions to the scalar  Klein-Gordon operator in the curved spacetime is an interesting and difficult problem in its own right. For  some  FLRW models such fundamental solutions were   recently written  via special integral transform in terms of the solution to the scalar wave equation in the Minkowski spacetime. For more details we refer the reader to  \cite{Yag_Galst_CMP,MN2015}.  

At the same time a straightforward approach to the Dirac operator  based on the separation of variables that was refraining from an explicit   factorization   led to some important sets of the exact solutions to the Dirac equation in FLRW spaces. (See, e.g., \cite{Barut-D,Oliveira,Finster,Huang_Justin,Villalba}.)  In particular, in \cite{Oliveira} are investigated the hydrogen atom, the Dirac-Morse oscillator, and the Dirac particle in a curved spacetime with the metric
\[
ds^2=e^{2f(r)}dt^2-e^{2g(r)}dr^2-r^2d\theta ^2-r^2\sin^2 \theta d\phi ^2. 
\]
 
 Since   time of publication of the articles by Fock \cite{Fock} and  
Schr\"odinger~\cite{Schrodinger}, many exact solutions of the Dirac equation    in the de~Sitter spacetime were obtained  (see, e.g., \cite{Barut-D,Huang_Justin,Huang,Zecca}  and bibliography therein). These exact solutions are products of  functions depending of single variable of time, radial and angular variables. They are produced by the separation of variables approach.
Separation of variables in the Dirac equation was
possible because of the simple form taken by the metric tensor in
FLRW models, and a selection of the comoving frame. 

The construction
of a quantum field theory in curved spacetimes
and the definition of a quantum vacuum demand  a detailed
investigation of the solutions of relativistic  equations
in curved backgrounds. (See, e.g., \cite{Birrell}.) The explicit formulas for all solutions and, in particular, for the fundamental solutions  of those equations may contribute in the resolving of that  challenging problem.

 Although the exact    solutions obtained by separation of variables approach answer to some very interesting   questions of physics (see, e.g., \cite{Barut-D,Brill-Wheeler,Finster,Huang_Justin,Huang2005}  and bibliography therein) in the de~Sitter spacetime, the explicit formulas for all  solutions   and the fundamental solutions were remaining open. Our paper fills that gap and presents the fundamental solutions and the solutions to the Cauchy problem via classical formulas for the wave equation in the Minkowski spacetime and the certain integral transform involving the Gauss's hypergeometric function in the kernel.  One can regard the integral transform as an analytical mechanism that from the massless field in the Minkowski spacetime generates massive particle in the curved spacetime.  As it is shown in Section~\ref{S2a} this mechanism exists even in the case of the vanishing cosmological constant when it provides spin zero and spin 1/2 particles with the mass due to massless scalar   field. We note that this mechanism appeals to some additional ``time'' variable.    
\medskip

  Even though nowadays, numerical solutions of   differential equations are 
available, in some cases    a deep understanding of 
 properties of the solutions  is  possible  only 
by the examination of the  explicit formulas when they are known. This is the case with the Huygens' principle. Some known results on the Huygens' principle
for the Dirac equation one can find in \cite{Deser-Nepomechie,Faraoni,Gunter,McLenaghan-Sasse,Pascazio,Sonego-Faraoni,Wunsch}. The application of the results of the present paper to the Huygens' principle will  be given in the forthcoming paper.

\bigskip

We start with the the fundamental solution to the Klein-Gordon operator in the FLRW model with the de~Sitter  metric.
 Recall  that   a retarded fundamental solution (a retarded inverse) for the Dirac operator (\ref{DE})  is a matrix operator  $ 
{\mathcal E}^{ret}={\mathcal E}^{ret} \left(x, t ; x_{0}, t_{0};m\right)
 $    that solves the equation
\begin{eqnarray}
\label{FSE} 
\left(  i {\gamma }^0    \partial_0   +i e^{-Ht}{\gamma }^\ell  \partial_\ell   +i     \frac{3}{2}   H  {\gamma }^0 -m{\mathbb I}_4 \right){\mathcal E}  \left(x, t ; x_{0}, t_{0};m\right)
& = &\delta\left(x-x_{0}, t-t_{0}\right){\mathbb I}_4, \nonumber \\
&  &
 (x,t,x_0,t_0 ) \in {\mathbb R}^8,  
\end{eqnarray}
and with the support in the {\it chronological future} (``forward light cone'') $D_+(x_0, t_0)$  of the point $(x_0,t_0)  \in {\mathbb R}^4$. The
advanced fundamental solution (propagator) $ 
{\mathcal E}^{adv}={\mathcal E}^{adv}  (x, t ; x_{0}, t_{0};$ $  m )
 $ solves the equation (\ref{FSE}) and has a  support in the {\it chronological past} (``backward light cone'') $D_-(x_0, t_0)$.  The forward and backward light cones are defined as follows:
\[ 
D_{\pm}\left(x_{0}, t_{0}\right) :=\left\{(x, t) \in {\mathbb R}^{3+1} ;
\left|x-x_{0}\right| \leq \pm\left(\phi (t) -\phi (t_{0})  \right)\right\}\,,
\]
where $\phi (t):= (1-e^{-Ht} )/H$ is a distance function.  
In fact, any intersection of $D_-(x_0, t_0)$ with the hyperplane $t = const < t_0$ determines the
so-called {\it dependence domain} for the point $ (x_0, t_0)$, while the intersection of $D_+(x_0, t_0)$
with the hyperplane $t = const > t_0$ is the so-called {\it domain of influence} of the point
$ (x_0, t_0)$. 
The Dirac equation (\ref{DE})  is non-invariant with respect to time inversion and its solutions have different properties in different direction of time.
\medskip

For the  construction of  the fundamental solutions of the Dirac equation we use the fundamental solutions of the Klein-Gordon equation in the de~Sitter spacetime from 
\cite{Yag_Galst_CMP,MN2015}. The Klein-Gordon  equation for the scalar field with the mass  $m$   
in the de~Sitter universe in the physical variables is:    
\[
\frac{1}{c^2}\psi _{tt} +   \frac{1}{c^2} 3 H   \psi _t - e^{-2tH} \Delta   \psi  + \ \frac{c^2 m^2}{h^2} \psi = 0\,.  
\]  
In fact,  
the function $u = e^{ \frac{3}{2}Ht}\psi $  solves the Klein-Gordon non-covariant equation  
\[
\frac{1}{c^2}u_{tt}     - e^{-2tH} \Delta   u - \left(\frac{9H^2}{4c^2}-\frac{c^2 m^2}{h^2}   \right)u=  0\,. 
\] 
For simplicity we set $c=1$ and $h=1$. 
We recall (see \cite{Yag_Galst_CMP,MN2015})  the fundamental solutions  of the Klein-Gordon equation in the de~Sitter spacetime. For $(x_0, t_0) \in {\mathbb R}^n\times {\mathbb R}$, $M \in {\mathbb C}$,  we define the function
\begin{eqnarray*}
E(x,t;x_0,t_0;M)
& :=  &
4^{-\frac{M}{H}} e^{ M  (  t_0+ t)} \left(\left(e^{-H t_0}+e^{-H t}\right)^2-(x - x_0)^2\right)^{\frac{M}{H}-\frac{1}{2}}  \\
&  &
\times  F \left(\frac{1}{2}-\frac{M}{H},\frac{1}{2}-\frac{M}{H};1;\frac{\left(e^{-H t}-e^{-H t_0}\right)^2-(x - x_0)^2}{\left(e^{-H t}+e^{-Ht_0}\right)^2-(x - x_0)^2}\right)\,, \nonumber
\end{eqnarray*}
where   $(x,t) \in D_+ (x_0,t_0)\cup D_- (x_0,t_0) $  and $F\big(a, b;c; \zeta \big) $ is the hypergeometric function (see, e.g.,\cite{B-E}). 
When no ambiguity arises,  we use the notation $x^2:= |x|^2$ for $x \in {\mathbb R}^n $.
Thus, the function $E  $ depends on $r^2= (x- x_0 )^2/H^2$, and we will write $E  (r,t;0,t_0;M) $ for   $E(x,t;x_0,t_0;M)  $:
\begin{eqnarray}
\label{0.6} 
E(r,t;0,t_0;M)
& :=  &
4^{-\frac{M}{H}} e^{ M  (  t_0+ t)} \left(\left(e^{-H t_0}+e^{-H t}\right)^2-(H r)^2\right)^{\frac{M}{H}-\frac{1}{2}}  \nonumber \\
&  &
\times  F \left(\frac{1}{2}-\frac{M}{H},\frac{1}{2}-\frac{M}{H};1;\frac{\left(-e^{-H t}+e^{-H t_0}\right)^2-(r H)^2}{\left(e^{-H t}+e^{-Ht_0}\right)^2-(r H)^2}\right)  \,.
\end{eqnarray}
Let $\Delta  $ be the Laplace operator in ${\mathbb R}^n $. For the Klein-Gordon non-covariant operator in the de~Sitter spacetime 
\begin{equation}
\label{KGO}
  {\mathcal S}_{ KG}=  \partial_t ^2  - e^{-2Ht} \Delta  -M^2  
\end{equation}
we define two  fundamental solutions \, ${\mathcal E}_{\pm,KG}(x,t;x_0,t_0;M) $ $(= {\mathcal E}_{\pm,KG}(x-x_0,t;0,t_0;M))$ as the  distributions ${\mathcal E}_{\pm,KG} \in {\mathcal  D}' ({\mathbb R}^{2n+2})$ with  
supports in the    cones  $D_\pm (x_0,t_0) $, $x_0 \in {\mathbb R}^n$, $t_0 \in {\mathbb R}$, \, 
supp$\,{\mathcal E}_{\pm,KG} \subseteq D_\pm (x_0,t_0)$, by
\[
\left(   \partial_t ^2  - e^{-2Ht}\bigtriangleup -M^2 \right){\mathcal E}_{\pm,KG}(x,t;x_0,t_0;M)= \delta (t-t_0)\delta (x-x_0)\,.
\]
 
Since all formulas for the contracting universe are evident modifications of ones for the expanding universe, in order to avoid unnecessary complications in the formulas, henceforth we restrict ourselves to the case of $H>0$.  According to    \cite{Yag_Galst_CMP,MN2015}, 
if $x \in {\mathbb R}^n$  and $M \in {\mathbb C}$, then for the operator $
{\mathcal S}_{ KG}$ (\ref{KGO}) 
the retarded  fundamental solution (retarded propagator) \, ${\mathcal E}_{+,KG}(x,t;x_0,t_0;M) $ $(= {\mathcal E}_{+,KG}(x-x_0,t;0,t_0;M))$ with  
support in the  forward cone  $D_+ (x_0,t_0) $, $x_0 \in {\mathbb R}^n$, $t_0 \in {\mathbb R}$, \, 
supp$\,{\mathcal E}_{+,KG} \subseteq D_+ (x_0,t_0)$, is given by the following integral  
\begin{eqnarray*}
{\mathcal E}_{+,KG}(x ,t;x_0 ,t_0;M)   
& = &
2   
  \int_{ 0}^{\phi (t)- \phi (t_0) } E  (r,t;0,t_0;M)    {\mathcal E}^w (x-x_0,r )  \, dr\,, \quad  t>t_0.
\end{eqnarray*}
Here  the distribution 
${\mathcal E}^w(x,t)$   
is a fundamental solution to the Cauchy problem for the  wave equation  in the Minkowski spacetime
\[
{\mathcal E}^w_{ tt} -   \bigtriangleup {\mathcal E}^w  =  0 \,, \quad {\mathcal E}^w(x,0)=\delta (x)\,, \quad {\mathcal E}^w_{t}(x,0)= 0\,.
\]
The  fundamental solution (advanced propagator) ${\mathcal E}_{-,KG}(x,t;x_0,t_0) $ $(= {\mathcal E}_{-,KG}(x-x_0,t;0,t_0))$ with  
support in the  backward cone   $D_- (x_0,t_0) $, $x_0 \in {\mathbb R}^n$, $t_0 \in {\mathbb R}$, \, 
supp$\,{\mathcal E}_{-,KG} \subseteq  D_- (x_0,t_0)$, is given by the following integral 
\begin{eqnarray*} 
{\mathcal E}_{-,KG}(x ,t;x_0,t_0;M) 
& = &
-2   
  \int_{\phi (t) -\phi (t_{0}) } ^{ 0}  E  (r,t;0,t_0;M)   {\mathcal E}^w (x-x_0,r ) \, dr\,, \quad  t<t_0 .
\end{eqnarray*} 
We remind that (see, e.g., \cite{Shatah})  if   $n$ is odd, then 
\[ 
{\mathcal E}^{w}(x, t) :=\frac{1}{\omega_{n-1} 1 \cdot 3 \cdot 5 \ldots(n-2)} \frac{\partial}{\partial t}\left(\frac{1}{t} \frac{\partial}{\partial t}\right)^{\frac{n-3}{2}} \frac{1}{t} \delta(|x|-t)\,,
\]
 while for the even  $n$ we have
\[ 
{\mathcal E}^{w}(x, t) \quad :=\frac{2}{\omega_{n-1} 1 \cdot 3 \cdot 5 \ldots(n-1)} \frac{\partial}{\partial t}\left(\frac{1}{t} \frac{\partial}{\partial t}\right)^{\frac{n-2}{2}} \frac{1}{\sqrt{t^{2}-|x|^{2}}} \chi_{B_{t}(x)}\,.
\]
 Here $\chi_{B_t(x)}$ denotes the characteristic function of the ball $B_{t}(x) :=\left\{x \in {\mathbb R}^{n} ;|x| \leq t\right\}$. The
constant $\omega _{n-1}$ is the area of the unit sphere $S^{n-1} \subset {\mathbb R}^{n}$. The distribution $\delta(|x|-t)$ is defined
 by
\[
\langle \delta(|\cdot|-t), \psi(\cdot)\rangle  =\int_{|x|=t} \psi(x) d x \quad \mbox{\rm for } \quad \psi \in C_{0}^{\infty}\left({\mathbb R}^{n}\right).
\]
\medskip

Denote  
\[
M_+=   \frac{1}{2}H  + im    ,\quad M_-=   \frac{1}{2}H  - im \,.
\]
Let  ${\mathcal E}_{ret,KG} $ be  a matrix with the operator-valued entries $\left( {\mathcal E}_{ret,KG} \left(x, t ; x_{0}, t_{0};m\right) \right)_{ij} $  defined  as follows 
\begin{eqnarray*}
   {\mathcal E}_{ret,KG} \left(x, t ; x_{0}, t_{0};m\right)  := 
\left( \begin{array}{cccc}
 {\mathcal E}_{+,KG}(x ,t;x_0 ,t_0;M_+) {\mathbb I}_2 & {\mathbb O}_2  \cr
{\mathbb O}_2 & {\mathcal E}_{+,KG}(x ,t;x_0 ,t_0;M_-){\mathbb I}_2 
\end{array} \right)\,.
\end{eqnarray*}
Let $e^{H\cdot } $  be  the operator multiplication by $e^{Ht } $. The next theorem  gives in the 
explicit form fundamental solutions of the the Dirac operator in the de~Sitter spacetime via the fundamental solutions of  
the wave operator in the Minkowski  spacetime.
\begin{theorem} 
\label{T0.1}
The fundamental solution $ 
{\mathcal E}^{ret}={\mathcal E}^{ret} \left(x, t ; x_{0}, t_{0};m\right)
 $ of the Dirac operator in the de~Sitter spacetime 
\begin{eqnarray*}
{\cal D} & = &
 i {\gamma }^0  \partial_0+i  e^{-Ht} {\gamma }^1\partial_1 +i e^{-Ht} {\gamma }^2 \partial_2 +i e^{-Ht}{\gamma }^3  \partial_3  + i       \frac{3}{2}  H  {\gamma }^0   -m{\mathbb I}_4\,, 
\end{eqnarray*}
$m \in {\mathbb C}$, is given by the following formula
\begin{eqnarray*} 
&  &
{\mathcal E}^{ret} \left(x, t ; x_{0}, t_{0};m\right) \\
& = &
- e^{-Ht}\left( i\gamma ^0 \partial_0+  ie^{-Ht} \gamma ^k \partial_k- i\frac{H}{2}\gamma ^0+m{\mathbb I}_4\right)  {\mathcal E}_{ret,KG}\left(x, t ; x_{0}, t_{0};m\right)[e^{H\cdot } ]\,.
 \end{eqnarray*}
\end{theorem}

The similar representation holds for the advanced propagator. 
The next theorem gives the representation formulas for the solutions of the Cauchy problem. We introduce the operator ${\cal G}(x,t,D_x;M ) $ by 
\begin{eqnarray*}
&  &
{\cal G}(x,t,D_x;M )[f] \\
&  =  &
2   \int_{ 0}^{t} db
  \int_{ 0}^{\phi (t)- \phi (b)}  E(r,t;0,b;M) \int_{{\mathbb R}^n} {\mathcal E}^w (x-y,r) f  (y,b) \,d y  \, dr  
, \quad  f \in C_0^\infty({\mathbb R}^{n+1})\,.
\end{eqnarray*}
Next,  
we define the kernel function
\begin{eqnarray*}
K_1(r,t;M)
& :=  &
E(r,t;0,0;M)   
\end{eqnarray*}
and the operator ${\cal K}_1(x,t,D_x;M)$   as follows:
\[
{\cal K}_1(x,t,D_x;M) \varphi (x) = 
 2\int_{0}^{\phi (t) } 
  K_1( s,t;M)  \int_{{\mathbb R}^n} {\mathcal E}^w (x-y,s)  \varphi  (y) \,d y\, ds\,, \quad \varphi \in C_0^\infty({\mathbb R}^n).
\]

\begin{theorem} 
\label{T0.2}
A solution to the Cauchy problem
\begin{eqnarray*}
&  &
\begin{cases} \dsp \left(i {\gamma }^0  \partial_0+i  e^{-Ht}{\gamma }^k \partial_k  + i       \frac{3}{2}  H  {\gamma }^0   -m{\mathbb I}_4\right)\Psi (x,t)=F(x,t)\,,\cr 
\Psi (x,0)= \Phi   (x ) \,,
\end{cases}
\end{eqnarray*}
$m \in {\mathbb C}$, is given by the following formula
\begin{eqnarray*} 
\Psi (x,t) 
& = &
 - e^{-Ht}\left( i\gamma ^0 \partial_0+  ie^{-Ht} \gamma ^k \partial_k- i\frac{H}{2}\gamma ^0+m{\mathbb I}_4\right) \\
&  &
\times \left [   \left (
   \begin{array}{cccc}
 {\cal G}(x,t,D_x;M_+){\mathbb I}_2&0   \\ 
 0 & {\cal G}(x,t,D_x;M_-){\mathbb I}_2   \\ 
   \end{array}
   \right )[e^{H\cdot }F ] \right.\\
&  &
\left. +i\gamma ^0 \left (
   \begin{array}{cccc}
 {\cal K}_1(x,t,D_x;M_+){\mathbb I}_2 &0  \\ 
0&  {\cal K}_1(x,t,D_x;M_-){\mathbb I}_2   \\ 
   \end{array}
   \right )[\Phi   ]\right ]  \,.
 \end{eqnarray*}
\end{theorem}

\begin{corollary}
In particular, for the massless particle (e.g., neutrino \cite{B-Sh,Brill-Wheeler, Huang_Justin}) $M_+=M_-=\frac{1}{2}H$ the last formula simplifies to the following one
\begin{eqnarray*} 
\Psi (x,t)  
& = &
 -  e^{-Ht}\left( i\gamma ^0 \partial_0+  ie^{-Ht} \gamma ^k \partial_k- i\frac{H}{2}\gamma ^0 \right) \\
&  &
\times 
  e^{\frac{H}{2}  t}\left [  \int_{ 0}^{t}
 \left(  \int_{ 0}^{\phi (t)- \phi (b)}    \int_{{\mathbb R}^n} {\mathcal E}^w (x-y,r) F  (y,b) \,d y  \, dr \right)  e^{\frac{3H}{2}  b} db  \right.\\
 &  &
\left. \hspace{4cm} +i\gamma ^0  
      \int_{0}^{\phi (t) } 
 \int_{{\mathbb R}^n} {\mathcal E}^w (x-y,r)    \Phi (y) \,d y\, dr    
    \right ]    \,.
 \end{eqnarray*}
\end{corollary}
In fact, for the massless  field  the kernel functions are simplified to 
\begin{equation*} 
E\left(r,t;0,t_0; \frac{H}{2}\right)
   =   
\frac{1}{2}  e^{\frac{H}{2}  (  t_0+ t)}\,,\quad  K_1\left(r,t;\frac{H}{2}\right)
=  \frac{1}{2}  e^{\frac{H}{2}  t}  
\end{equation*}
and that proves the corollary.
 \medskip

\medskip

The paper is organized as follows. In Section~\ref{S1} we introduce the  formulas for the solution of the generalized Klein-Gordon equation in the de~Sitter and Minkowski spacetimes. In Section~\ref{S2} we introduce the generalized Dirac operator and show how the  generalized Klein-Gordon  operator in the de~Sitter and Minkowski spacetimes can be factorized into a product of two first-order matrix  coefficients operators with the generalized Dirac operator in the  same spacetimes as the first factor. Examples of such factorizations are also provided in that section.  Section~\ref{S3} is devoted to the completion of the proof of Theorem~\ref{T0.1}.

\smallskip

\section{Representation of the Solution of the Dirac equation in de~Sitter spacetime} 
\label{S1}

Additional to (\ref{0.6}) 
for $M \in {\mathbb C}$  we recall two more kernel functions from \cite{Yag_Galst_CMP,MN2015}
\begin{eqnarray}
\label{K0MH}
K_0(r,t;M)
&  := &
-\left[\frac{\partial}{\partial b} E(r, t ; 0, b ; M)\right]_{b=0}\,,   \\
\label{K1MH}
K_1(r,t;M)
& :=  &
E(r,t;0, 0;M)   \,.  
\end{eqnarray}
Then according to \cite{MN2015} the solution operator for the Cauchy problem for the scalar {\it generalized Klein-Gordon equation}
in the de~Sitter spacetime  
\begin{eqnarray}
\label{6.17}
&  &
\left( \partial_0^2     
-  e^{-2Ht}    {\mathcal A}(x,\partial_x)     
 -  M ^2   \right)\psi  =f,\quad \psi (x,0)= \varphi _0 (x), \quad \psi_t (x,0)= \varphi _0 (x) \,,  
 \end{eqnarray}
is given as follows
\[
\psi (x,t) = {\cal G} ( x,t,D_x;M)[f]+{\cal K}_0(x,t,D_x;M)[\varphi _0 ]+ {\cal K}_1(x,t,D_x;M)[\varphi _1 ] \,.
\]
Here ${\mathcal A}(x,\partial_x) $ is the   differential operator ${\mathcal A}(x,\partial_x)=\sum_{|\alpha| \leq m} a_\alpha (x)D_x^\alpha $ and the coefficients $ a_\alpha (x)$ 
are $C^\infty$-functions in the open domain $\Omega \subseteq {\mathbb R}^n $, that is $a_\alpha \in C^\infty (\Omega ) $. 
The kernels   $K_0(z,t;M) $   and  $K_1(z,t;M) $ can be written in the explicit form as follows
\begin{eqnarray*}
K_0(r,t;M) 
\!&\!=\!&\!
-4^{-\frac{M}{H}} \left((1+e^{  -H t})^2- H^2 r^2\right)^{\frac{M}{H}-\frac{5}{2}} e^{t (-4 H+M)}
 \\
&  &
\times \Bigg\{ \! e^{  -2H t}\left( (1+e^{  -H t})^2- H^2 r^2  \right) \!\!\left(-e^{2 H t} \left(H  (H M r^2-1 )+M\right)+H e^{H t}+M\right)  \\
&  &
\times F \left(\frac{1}{2}-\frac{M}{H},\frac{1}{2}-\frac{M}{H};1;\frac{\left(1-e^{-H t}\right)^2-H^2 r^2}{\left(1+e^{-H t}\right)^2-H^2 r^2}\right)\\
&  &
+\frac{1}{H} (H-2 M)^2 e^{3t H}   \left( e^{-2 H t}-\left(H^2 r^2+1\right) \right)  
 \\
&  &
\left. \times F \left(\frac{3}{2}-\frac{M}{H},\frac{3}{2}-\frac{M}{H};2;\frac{\left(1-e^{-H t}\right)^2-H^2 r^2}{\left(1+e^{-H t}\right)^2-H^2 r^2}\right)\right\},\\ 
K_1(r,t;M)
& =  &
4^{-\frac{M}{H}} e^{M t} \left(\left(1+e^{-H t}\right)^2-(H r)^2\right)^{\frac{M}{H}-\frac{1}{2}} \\
&  &
\times  F \left(\frac{1}{2}-\frac{M}{H},\frac{1}{2}-\frac{M}{H};1;\frac{\left(1-e^{-H t} \right)^2-(r H)^2}{\left(1+e^{-H t} \right)^2-(r H)^2}\right) \nonumber \,.  
\end{eqnarray*}

To describe the operators ${\mathcal G}, {\mathcal K}_1 $ we 
  recall the results of  Theorem~1.1~\cite{MN2015}.  
For $ f \in C (\Omega\times I  )$,\, $ I=[0,T]$, $0< T \leq \infty$, and \, $ \varphi_0 $,  $ \varphi_1 \in C (\Omega ) $,
let  the function\,
$v_f(x,t;b) \in C_{x,t,b}^{m,2,0}(\Omega \times [0,(1-e^{-HT})/H]\times I)$\,
be a solution to the   problem
\begin{equation}
\label{1.22}
\begin{cases} 
 v_{tt} -   {\mathcal A}(x,\partial_x)  v  =  0 \,, \quad x \in \Omega \,,\quad t \in [0,(1-e^{-HT})/H]\,,\cr
v(x,0;b)=f(x,b)\,, \quad v_t(x,0;b)= 0\,, \quad b \in I,\quad x \in \Omega\,, 
\end{cases}
\end{equation}
and the function \, $  v_\varphi(x, t) \in C_{x,t}^{m,2}(\Omega \times [0,(1-e^{-HT})/H])$ \, be a   solution   of the   problem
\begin{equation}
\label{1.23}
\begin{cases} 
 v_{tt}-   {\mathcal A}(x,\partial_x)  v =0, \quad x \in \Omega \,,\quad t \in [0,(1-e^{-HT})/H]\,, \cr
 v(x,0)= \varphi (x), \quad v_t(x,0)=0\,,\quad x \in \Omega\,.  
\end{cases} 
\end{equation}
Then the function  $u= u(x,t)$   defined by
\begin{eqnarray*}
u(x,t)
&  =  &
2   \int_{ 0}^{t} db
  \int_{ 0}^{\phi (t)- \phi (b)}  E(r,t;0,b;M)  v_f(x,r ;b) \, dr  
+ e ^{\frac{Ht}{2}} v_{\varphi_0}  (x, \phi (t))\\
&  &
+ \, 2\int_{ 0}^{\phi (t)}  K_0( s,t;M)v_{\varphi_0}  (x, s)   ds  \nonumber 
+\, 2\int_{0}^{\phi (t) }  v_{\varphi _1 } (x,  s)
  K_1( s,t;M)   ds
, \quad x \in \Omega  , \,  t \in I ,
\end{eqnarray*}
where $\phi (t):= (1-e^{-Ht} )/H$,
 solves the problem
\[
\begin{cases}
u_{tt} - e^{-2Ht}{\mathcal A}(x,\partial_x)  u - M^2 u= f, \quad  x \in \Omega \,,\,\, t \in I,\cr
  u(x,0)= \varphi_0 (x)\, , \quad u_t(x,0)=\varphi_1 (x),\quad x \in \Omega\,.
  \end{cases}
\]
Here the kernels  $E$, $K_0$ and $K_1$ have been defined in (\ref{0.6}), (\ref{K0MH}) and (\ref{K1MH}), respectively.

\begin{remark}
We stress here that the existence of the solutions in the problems (\ref{1.22}) and (\ref{1.23})  is assumed.
For the case of second-order elliptic operator ${\mathcal A} (x,\partial_x) $   in order to guarantee existence of the solutions in the problems (\ref{1.22}) and (\ref{1.23})   the initial data must be given in the  domain $ \{x\in {\mathbb R}^n | dist(x,\Omega ) \leq  c \}  $, where  $c>0$ due to the propagation phenomena.  
\end{remark}

Moreover, if ${\mathcal E}^w_{\mathcal A}(x ,t;x_0 ) $ is a fundamental solution of the Cauchy problem for the massless equation (\ref{1.23}) in the spacetime without expansion, that is,  
\[ 
\begin{cases}
\left( \partial_{t}^2 - {\mathcal A}(x,\partial_x)\right){\mathcal E}^w_{\mathcal A}(x ,t;x_0 )  =0, \quad  x \in {\mathbb R}^n \,,\,\, t \in I,\cr
 {\mathcal E}^w_{\mathcal A} (x,0;x_0)= \delta (x-x_0)\, , \quad \partial_t{\mathcal E}^w_{\mathcal A} (x,0;x_0) =0,\quad x \in  {\mathbb R}^n\,,
  \end{cases}
\]
then, the fundamental solutions ${\mathcal E}_{\pm,KG,{\mathcal A}}$ of the operator $\partial_{t}^2  - e^{-2Ht}{\mathcal A}(x,\partial_x)- M^2 $ can be written as follows 
\[  
{\mathcal E}_{\pm,KG,{\mathcal A}}(x ,t;x_0 ,t_0;M)   
  =  
2   
  \int_{ 0}^{\phi (t)- \phi (t_0) } E  (r,t;0,t_0;M)   {\mathcal E}^w_{\mathcal A} (x,r ;x_0)  \, dr,\quad \text{when} \,\,\pm (t-t_0)>0 
\]
on their supports.

We mention here two important examples. The first one has ${\mathcal A}(x,\partial_x) =\Delta  $ and it is related to the problem written in the Cartesian coordinates. The second one has the equation written in the spherical coordinates $( r ,   \theta, \phi)  $. In the  FLRW  spacetime with the  line element
\begin{equation}
\label{dScurved}
ds^2=    dt^2
- e^{2Ht}\left( \frac{1}{1-Kr^2} dr^2 + r^2(d\theta ^2 + \sin^2 \theta \, d\phi ^2) \right)
\end{equation} 
the Klein-Gordon equation is
\begin{eqnarray}
\label{1.9nn}
&   &
 \partial_t ^2    \psi
+3 H\partial_ t \psi
  -  e^{-2Ht}{\mathcal A}(x,\partial_x)  \psi  + m^2 \psi =0\,,
\end{eqnarray}
where  
\begin{eqnarray*}
{\mathcal A}(x,\partial_x) 
&  :=   &
\frac{\sqrt{1-Kr^2}}{ r^2 }\frac{\partial }{\partial r}\left(  r^2  \sqrt{1-Kr^2}\frac{\partial \psi }{\partial  r} \right)\\
&  &
+   \frac{1}{ r^2\sin  \theta}\frac{\partial }{\partial \theta }\left(    \sin  \theta  \frac{\partial \psi }{\partial \theta } \right)
+   \frac{1 }{ r^2\sin^2  \theta}\frac{\partial }{\partial \phi }\left( \frac{\partial \psi }{\partial \phi  } \right)
\end{eqnarray*}
is the Laplace-Beltrami operator in the  spatial variables, where $K=-1, 0$, or $+1$, for a hyperbolic, flat, or spherical spatial geometry, respectively. In the Cartesian coordinates in the metric (\ref{dScurved}) the covariant Klein-Gordon  equation    reads  (\ref{1.9nn}), where (see \cite[Example 2]{Galstian-Yagdjian2020})
\begin{eqnarray*}
{\mathcal A}(x, \partial_x )
& = &
( 1-Kx_1^2) \partial^2 _{x_1}  +(1-Kx_2^2)\partial_{x_2}^2   
+(1-Kx_3^2)\partial_{x_3}^2   -2 Kx_1 x_2 \partial_{x_1}\partial_{x_2} \\
&  &  
-2 Kx_1 x_3  \partial_{x_1}\partial_{x_3}   
-2K x_2 x_3 \partial_{x_2}\partial_{x_3}   
-3K x_1  \partial_{x_1} 
-3K x_2  \partial_{x_2}  
-3K x_3  \partial_{x_3} \\
& = &
\Delta -K(x_1\partial_{x_1}+x_2\partial_{x_2}+x_3\partial_{x_3})^2-K(x_1\partial_{x_1}+x_2\partial_{x_2}+x_3\partial_{x_3})
\end{eqnarray*}
and $ x_1\partial_{x_1}+x_2\partial_{x_2}+x_3\partial_{x_3}$ is the radial vector field. 
\medskip

We note here that the transition from the Laplace operator in ${\mathbb R}^n$ to more general operator ${\mathcal A}(x,\partial_x)  $ allows us to consider more general equations 
than ones  generated by the change of coordinates. 
\medskip

The function $u = e^{\frac{3}{2}Ht}\psi $  solves the generalized scalar Klein-Gordon equation  
\begin{eqnarray*}
&   &
 \partial_t ^2  u 
  -  e^{-2t}{\mathcal A}(x,\partial_x)  u  + \left( m^2 -\frac{9}{4}H^2 \right)u =0\,.
\end{eqnarray*}
The next theorem gives a representation of the solution of the Cauchy problem for the $4\times 4$ system of the 
generalized non-covariant Klein-Gordon equations with complex valued mass in the de~Sitter spacetime.  
\begin{theorem}
\label{T1.1}
Let a vector-valued function $\Psi $ be a solution to the equation   
\begin{eqnarray}
\label{Nc_KG_system}
&  &
\left( \partial_0^2     
-  e^{-2Ht}    {\mathcal A}(x,\partial_x)  \right){\mathbb I}_4\Psi    
+ \left( m {\mathbb I}_4 -  i\frac{H}{2}\gamma ^0\right) ^2  \Psi   =F \,.
 \end{eqnarray}
Then its components $\Psi_0(x,t),\Psi_1(x,t),\Psi_2(x,t),\Psi_3(x,t)  $ can be written as follows: for $i=0,1$
 \[
\Psi_i(x,t) ={\cal G} (x,t,D_x;M_+)[F_i]+{\cal K}_0(x,t,D_x;M_+)[\Psi_i(x,0) ]+ {\cal K}_1(x,t,D_x;M_+)[\partial_0\Psi_i(x,0) ],  
 \]
while for $k=2,3$
 \[
\Psi_k(x,t) ={\cal G} (x,t,D_x;M_-)[F_k]+{\cal K}_0(x,t,D_x;M_-)[\Psi_k(x,0) ]+ {\cal K}_1(x,t,D_x;M_-)[\partial_0\Psi_k(x,0)] .
 \]
 \end{theorem}
\medskip

\noindent
{\bf Proof.} Equation (\ref{Nc_KG_system}) can be written as follows
\begin{eqnarray*}
&  &
\left( \partial_0^2     
-  e^{-2Ht}    {\mathcal A}(x,\partial_x)      
+ m ^2  -  \frac{1}{4}H^2 \right){\mathbb I}_4\Psi -imH\gamma ^0  \Psi =F .
 \end{eqnarray*} 
We write the last equation in the components  $\Psi =(\Psi _0,\Psi _1,\Psi _2,\Psi _3)^T $ ($A^T$ means transposition of the matrix $A$) as follows
\begin{eqnarray*}
&  &
\left( \partial_0^2     
-  e^{-2Ht}    {\mathcal A}(x,\partial_x)      
 + m ^2 -  \frac{1}{4}H^2-imH \right)\Psi_j =F_j,\quad j=0,1, \\
&  &
\left( \partial_0^2     
-  e^{-2Ht}  {\mathcal A}(x,\partial_x)      
 + m ^2 -  \frac{1}{4}H^2+imH \right)\Psi_k =F_k,\quad k=2,3\,.
 \end{eqnarray*}
 It remains to apply Theorem~1.1~\cite{MN2015}. Theorem is proved.
\qed
\smallskip

The last theorem can be easily extended to the diagonal system of Klein-Gordon equations 
\begin{eqnarray}
\label{Nc_KGM_system}
&  &
\left( \partial_0^2     
-  e^{-2Ht}    {\mathcal A}(x,\partial_x)  \right){\mathbb I}_4\Psi    
+ {\mathcal M}  \Psi   =F \,,
 \end{eqnarray}
where $ {\mathcal A}(x,\partial_x)$ is a diagonal matrix of operators, while ${\mathcal M} $ is a  constant diagonal matrix ${\mathcal M}=diag(M_{i } ) $, $i=1,\ldots,m$, with the entries $ M_{i } \in {\mathbb C}  $
possibly depending on some parameters. Indeed, if we introduce the diagonal matrices of the operators 
 \begin{eqnarray*}
&  &
({\bf \cal G} (x,t,D_x;{\mathcal M} ))_{jk}=\delta_{jk} {\cal G} (x,t,D_x;M_j),\\ 
&  &
({\cal K}_0(x,t,D_x;{\mathcal M}))_{jk}=\delta_{jk} {\cal K}_0(x,t,D_x;M_j),\quad 
({\cal K}_1(x,t,D_x;{\mathcal M}))_{jk}=\delta_{jk} {\cal K}_1(x,t,D_x;M_j)\,,
\end{eqnarray*}
then the solution to the equation (\ref{Nc_KGM_system}) is given by 
 \begin{eqnarray*}
&  &
\Psi (x,t) ={\cal G} (x,t,D_x;{\mathcal M})[F ]+{\cal K}_0(x,t,D_x;{\mathcal M})[\Psi (x,0) ]+ {\cal K}_1(x,t,D_x;{\mathcal M})[\partial_0\Psi (x,0) ].  
 \end{eqnarray*}

Theorem~\ref{T1.1} allows us to write solution of the  {\it generalized Dirac equation} in the de~Sitter spacetime. In order to define   that class of Dirac operators we note that the Dirac  operator in the Minkowski spacetime  written in the Cartesian or curvilinear  coordinates is a member  of the family of operators of   the form 
\[
 i {\gamma }^0  \partial_0+i   \gamma ^k A_k(x, \partial_x)     -m{\mathbb I}_4   , 
 \]
where the operators $A_k(x, \partial_x) = \sum_{\ell=1}^3a_{k\ell}(x)\partial_{x_\ell}$, $k=1,2,3$, have variable coefficients depending on the spatial variables. Having in mind 
the diagonal system of Klein-Gordon equations (\ref{Nc_KGM_system}) and  four diagonal   linearly independent matrices  
\[
{\mathbb I}_4 ,\quad {\gamma }^0,\quad \gamma ^1\gamma ^2=-i\left(
\begin{array}{cccc}
 \sigma ^3 & {\mathbb O}_2   \\
 {\mathbb O}_2  & \sigma ^3   \\ 
\end{array}
\right),\quad -i\gamma ^3\gamma ^0\gamma ^1\gamma ^2\gamma ^3=\left(
\begin{array}{cccc}
- \sigma ^3 & {\mathbb O}_2    \\
 {\mathbb O}_2  & \sigma ^3   \\ 
\end{array}
\right),
\]
we can then define  {\it generalized Dirac operator} 
\[
i {\gamma }^0  \partial_0+i   \gamma ^k A_k(x, \partial_x)     -m{\mathbb I}_4 
\]
 by the 
condition 
\begin{equation}
\label{CAA}
\gamma ^k  A_k(x, \partial_x)  \gamma ^j A_j(x, \partial_x)= -   {\mathcal A} (x, \partial_x){\mathbb I}_4+{\mathcal B} (x, \partial_x){\gamma }^0+{\mathcal C} (x, \partial_x)\gamma ^1\gamma ^2+{\mathcal D} (x, \partial_x)\gamma ^3\gamma ^0\gamma ^1\gamma ^2\gamma ^3 
\end{equation}
imposed on,  in general,  the scalar pseudo-differential operators  $A_k(x, \partial_x)$, $ {\mathcal A} (x, \partial_x)$, ${\mathcal B} (x, \partial_x) $, ${\mathcal C} (x, \partial_x) $,  and ${\mathcal D} (x, \partial_x) $.  For the purpose of this paper it suffices to consider the case of ${\mathcal D} (x, \partial_x)=0$. Accordingly, we define {\it generalized Dirac operator
in the de~Sitter spacetime}
\[
i {\gamma }^0  \partial_0+i   e^{-Ht} \gamma ^k A_k(x, \partial_x)  +i   \frac{3}{2}  H    \gamma ^0   -m{\mathbb I}_4 \,,
\]
where 
${\mathcal A} (x, \partial_x)$, $A_k(x, \partial_x) $, $k=1,2,3$, are the scalar operators with  the property 
\begin{eqnarray}
\label{AA}
&  &
\gamma ^k  A_k(x, \partial_x)  \gamma ^j A_j(x, \partial_x)= -   {\mathcal A} (x, \partial_x){\mathbb I}_4\,.
\end{eqnarray}   
 Theorem~\ref{T0.2} is a particular case of the next theorem.
\begin{theorem} 
\label{T1.2}
Assume that ${\mathcal A} (x, \partial_x)$, $A_k(x, \partial_x) $, $k=1,2,3$, are the scalar operators with  the properties (\ref{AA}). 
Then the solution to the Cauchy problem
\begin{eqnarray}
\label{1.9n}
&  &
\begin{cases} \dsp \left(i {\gamma }^0  \partial_0+i e^{-Ht} \gamma ^k A_k(x, \partial_x)  + i       \frac{3}{2}  H  {\gamma }^0   -m{\mathbb I}_4\right)\Psi (x,t)=F(x,t),\cr 
\Psi (x,0)= \Phi   (x ) 
\end{cases}
\end{eqnarray}
is given as follows
\begin{eqnarray*} 
\Psi (x,t) 
& = &
- e^{-Ht}\left( i\gamma ^0 \partial_0+  ie^{-Ht} \gamma ^k A_k(x, \partial_x)- i\frac{H}{2}\gamma ^0+m{\mathbb I}_4\right) \\
&  &
\times \left [  \left (
   \begin{array}{cccc}
 {\cal G}(x,t,D_x;M_+){\mathbb I}_2 & {\mathbb O}_2  \\ 
{\mathbb O}_2 &  {\cal G}(x,t,D_x;M_-){\mathbb I}_2    \\ 
   \end{array}
   \right )[e^{H\cdot }F ] \right.\\
&  &
\left.+i\gamma ^0 \left (
   \begin{array}{cccc}
 {\cal K}_1(x,t,D_x;M_+) {\mathbb I}_2&{\mathbb O}_2  \\
 {\mathbb O}_2 & {\cal K}_1(x,t,D_x;M_-) {\mathbb I}_2 \\
   \end{array}
   \right )[\Phi  (x ) ] \right ]  \,.
 \end{eqnarray*}
\end{theorem}
\medskip

\noindent
{\bf Proof.} Indeed we can write
\begin{eqnarray} 
\label{2.7}
\Psi (x,t) 
& = &
 e^{-Ht}\left( i\gamma ^0 \partial_0+  ie^{-Ht} \gamma ^k A_k(x, \partial_x)- i\frac{H}{2}\gamma ^0+m{\mathbb I}_4\right) X\,,
 \end{eqnarray}
 where 
\begin{eqnarray*} 
X & = &
-\left (
   \begin{array}{cccc}
 {\cal G} (x,t,D_x;M_+) {\mathbb I}_2&{\mathbb O}_2   \\ 
 {\mathbb O}_2  & {\cal G}(x,t,D_x;M_-){\mathbb I}_2   \\ 
   \end{array}
   \right ) [ e^{H\cdot }F ] \\
&  &
 +  \left (
   \begin{array}{cccc}
   {\cal K}_1(x,t,D_x;M_+) {\mathbb I}_2 &{\mathbb O}_2  \\ 
 {\mathbb O}_2   &  {\cal K}_1(x,t,D_x;M_-) {\mathbb I}_2   \\ 
   \end{array}
   \right )  [(\partial_0 X ) (x,0) ]
\end{eqnarray*}
 and
 \[
 X(x,0)=0, \quad (\partial_0 X)(x,0)= -i\gamma ^0 \Phi (x)\,.
 \]
Then  the substitution of (\ref{2.7}) into the equation of (\ref{1.9n})  and 
 the relation (\ref{AA})  together with Proposition~\ref{P3.1} imply
\begin{eqnarray*} 
  &  &
\left( \left( \partial_0^2     
-  e^{- 2Ht}   {\mathcal A} (x, \partial_x)    
+ m ^2  -  \frac{1}{4}H^2 \right){\mathbb I}_4   +imH\gamma ^0  \right) X  =-e^{Ht} F\,. 
\end{eqnarray*} 
Consequently,
\begin{eqnarray*} 
&  &
 \left( i\gamma ^0 \partial_0+  ie^{-Ht} \gamma ^k A_k(x, \partial_x)+ i\frac{3H}{2}\gamma ^0-m{\mathbb I}_4\right) 
\Psi (x,t)=  \\ 
& = &
 \left( i\gamma ^0 \partial_0+  ie^{-Ht} \gamma ^k A_k(x, \partial_x)+ i\frac{3H}{2}\gamma ^0-m{\mathbb I}_4\right)\\
&  &
\times  
 e^{-Ht}\left( i\gamma ^0 \partial_0+  ie^{-Ht} \gamma ^k A_k(x, \partial_x)- i\frac{H}{2}\gamma ^0+m{\mathbb I}_4\right) X =\\ 
& = &
- e^{-Ht} \left( \partial_0^2     
-  e^{- 2Ht}  {\mathcal A} (x, \partial_x)   
+ m ^2  -  \frac{1}{4}H^2 +imH\gamma ^0 \right)     X \\
& = &
F\,.
\end{eqnarray*}
For the initial value of $\Psi (x,0) $ we have
\begin{eqnarray*} 
\Psi (x,0) 
& = &
 \lim _{t\to 0}e^{- Ht} \left( i\gamma ^0 \partial_0+  ie^{-Ht} \gamma ^k A_k(x, \partial_x)- i\frac{H}{2}\gamma ^0+m{\mathbb I}_4\right) X\\
& = &
   ( i\gamma ^0  ) (-i\gamma ^0 )\Phi (x)=\Phi (x)\,.
 \end{eqnarray*}
Theorem is proved. \qed
\smallskip

In Example 5 of Section~\ref{S2} we can apply Theorem~\ref{T1.2} to Dirac field in the de~Sitter spacetime in the presence of constant magnetic field by using the  further  generalization of the relation (\ref{CAA}).  

\smallskip

\section{The case of vanishing $H$}
\label{S2a}

Now we turn to the limit case of  Theorem~\ref{T1.2} as $H  $ approaches zero. This includes, in particular,   the Dirac equation in the Minkowski spacetime. That theorems shows how massless field via the integral transform provides mass to the massive field. In fact, the integral transform approach leads to the   formulas for the solutions also of the generalized Dirac equation. We start with the Klein-Gordon equation by the following theorem. We skip a proof that it can be done  by  straightforward calculations.

\begin{theorem}
The function  $u=u (x,t)  $   defined by
\begin{eqnarray*}
u(x,t)
&  =  &
\int_{ t_0}^{t} db
  \int_{ 0}^{t-b }   I_0\left(M\sqrt{(t-b)^2-r^2} \right) v_f(x,r ;b) \, dr
\\
&  &
+ v_{\varphi_0}  (x, t)  -  \int_{0 }^{t   }  \frac{iMt}{\sqrt{t^2-r^2} } J_1\left(iM\sqrt{t^2-r^2} \right)  v_{\varphi_0}  (x, r)  \, dr   \nonumber \\
&  &
+\,    \int_{0  }^{t }   I_0\left(M\sqrt{t^2-r^2} \right)   v_{\varphi_1}  (x, r) \, dr
, \quad x \in \Omega  , \,\, t \in [0,T]\,,
\end{eqnarray*}
where $ M \in {\mathbb C}$,   $v_f(x,r ;b) $ is a solution of  (\ref{1.22}), while $v_{\varphi_0}  (x, s)  $ and $v_{\varphi_1}  (x, s)  $ solve  the problem (\ref{1.23}),  is a solution of  the problem
\begin{eqnarray}
\label{1.20}
\begin{cases}
u_{tt} - {\mathcal A}(x,\partial_x)  u - M^2 u= f, \quad  x \in \Omega \,,\,\, t \in [0,T],\cr
  u(x,0)= \varphi_0 (x)\, , \quad u_t(x,0)=\varphi_1 (x),\quad x \in \Omega\,.
  \end{cases}
\end{eqnarray}
Here $I_\nu  $ and $J_\nu  $ are  the Bessel functions. 
\end{theorem}

Moreover, similar statement true for the fundamental solution of the generalized Klein-Gordon operator. We skip the proof of the next statement since it similar  to one of the previous theorem.  
\begin{theorem}
If ${\mathcal E}^w_{\mathcal A}={\mathcal E}^w_{\mathcal A}(x,t;x_0) $ is a fundamental solution of the Cauchy problem for the massless equation 
that is,  
\[ 
\begin{cases}
\left( \partial_{t}^2 - {\mathcal A}(x,\partial_x)\right){\mathcal E}^w_{\mathcal A} (x,t;x_0) =0, \quad  x,x_0 \in {\mathbb R}^n \,,\,\, t \in {\mathbb R},\cr
 {\mathcal E}^w_{\mathcal A} (x,0;x_0)= \delta (x-x_0)\, , \quad  \partial_t{\mathcal E}^w_{\mathcal A} (x,0;x_0) =0,\quad x,x_0 \in  {\mathbb R}^n\,,
  \end{cases}
\]
then, the fundamental solutions ${\mathcal E}_{\pm,KG,{\mathcal A}}$ of the operator of (\ref{1.20})   can be written as follows 
\[  
{\mathcal E}_{\pm,KG,{\mathcal A}}(x ,t;x_0 ,t_0;M)   
  =  
2   
  \int_{ 0}^{t- t_0 } I_0\left(M\sqrt{(t-t_0)^2-r^2} \right)   {\mathcal E}^w_{\mathcal A} (x,r; x_0)  \, dr,\quad \text{when} \,\,\pm (t-t_0)>0 
\]
on their supports.
\end{theorem}

The fundamental solutions ${\mathcal E}^w_{\mathcal A}={\mathcal E}^w_{\mathcal A}(x,t;x_0) $ for a wide class of hyperbolic operators are constructed in \cite[Ch.1,Ch.4]{Yagbook} (see also referenced therein).  The application of the previous two theorems leads to the following result. Define the operators ${\mathcal G}  $ and $ {\mathcal K}$   as follows
 \begin{eqnarray*} 
 {\cal G}^{Min}(x,t,D_x;M)[ f]
 &:=&
 \int_{ t_0}^{t} db
  \int_{ 0}^{t-b }   I_0\left(M\sqrt{(t-b)^2-r^2} \right) v_f(x,r ;b) \, dr\,,
 \\
{\cal K}^{Min}_1(x,t,D_x;M)[\varphi (x ) ] 
 &:=&
  \int_{0  }^{t }   I_0\left(M\sqrt{t^2-r^2} \right)   v_{\varphi }  (x, r) \, dr
, \quad x \in \Omega  , \,\, t \in {\mathbb R}_+\,, \\
 \end{eqnarray*}
 while
  $v_f(x,r ;b) $ is a solution of  
\[
\begin{cases} 
 v_{tt} -   {\mathcal A}(x,\partial_x)  v  =  0 \,, \quad x \in \Omega \,,\quad t \in {\mathbb R}_+\,,\cr
v(x,0;b)=f(x,b)\,, \quad v_t(x,0;b)= 0\,, \quad b \in {\mathbb R}_+,\quad x \in \Omega\,, 
\end{cases}
\] 
 and    $v_{\varphi }  (x, s)  $   solves  the problem   
\[
\begin{cases} 
 v_{tt}-   {\mathcal A}(x,\partial_x)  v =0, \quad x \in \Omega \,,\quad t \in {\mathbb R}_+\,, \cr
 v(x,0)= \varphi (x), \quad v_t(x,0)=0\,,\quad x \in \Omega\,.  
\end{cases} 
\]
\begin{theorem}
Assume that ${\mathcal A} (x, \partial_x)$, $A_k(x, \partial_x) $, $k=1,2,3$, are the scalar operators with  the properties
(\ref{AA}). 
Then the solution $\Psi =\Psi (x,t) $ to the Cauchy problem
\begin{eqnarray*}
&  &
\begin{cases} \dsp \left(i {\gamma }^0  \partial_0+i  \gamma ^k A_k(x, \partial_x)    -m{\mathbb I}_4\right)\Psi (x,t)=F(x,t), \quad x \in \Omega  , \,\, t \in {\mathbb R}_+\,,\cr 
\Psi (x,0)= \Phi   (x ), \quad x \in \Omega   \,, 
\end{cases}
\end{eqnarray*}
is given as follows
\begin{eqnarray*} 
&  &
\Psi (x,t) \\
& = &
-\left( i\gamma ^0 \partial_0+  i  \gamma ^k A_k(x, \partial_x)+m{\mathbb I}_4\right)  \left(  \left (
   \begin{array}{cccc}
 {\cal G}^{Min}(x,t,D_x;im){\mathbb I}_2 &{\mathbb O}_2  \\ 
 {\mathbb O}_2 & {\cal G}^{Min}(x,t,D_x;-im) {\mathbb I}_2  \\ 
   \end{array}
   \right )[ F ]\right.\\
&  &
\left. +i\gamma ^0 \left (
   \begin{array}{cccc}
 {\cal K}^{Min}_1(x,t,D_x;im) {\mathbb I}_2 &{\mathbb O}_2 \\ 
{\mathbb O}_2 &  {\cal K}^{Min}_1(x,t,D_x;-im) {\mathbb I}_2  \\ 
   \end{array}
   \right )[\Phi  ]\right )  \,.
 \end{eqnarray*}

\end{theorem}

\section{Factorization of the generalized Klein-Gordon operator in \\de~Sitter spacetime}
\label{S2}

In this section we introduce some factorization of the  generalized Klein-Gordon operator in de~Sitter spacetime with scalar diagonal  mass matrix. We believe that that  factorization is interesting in its own right and therefore we illustrate it with several examples.

In the next proposition, by appealing to  the complex-valued  mass matrix, we involve the Dirac operator in the de~Sitter spacetime in the factorization of the  generalized Klein-Gordon operator. In view of Pauli's fundamental theorem  \cite{Pauli} for the given generalized Klein-Gordon operator the next factorization is not unique.

\begin{proposition}
\label{P3.1}
Let ${\mathcal A} (x, \partial_x)$, $A_k(x, \partial_x) $, $k=1,2,3$, be  scalar operators with  the property (\ref{AA}) 
and $a,b,c \in{\mathbb C}$.   Then
\begin{eqnarray*}
&  &
e^{aHt}\left(i \gamma ^0 \partial_0+ i e^{-Ht} \gamma ^k A_k(x, \partial_x) + ib\frac{H}{2}  \gamma ^0 -m {\mathbb I}_4\right)\\
&  &
\times e^{-aHt}\left(i \gamma ^0 \partial_0+ i e^{-Ht} \gamma ^j A_j(x, \partial_x)-i(b-c)\frac{H}{2}  \gamma ^0+m {\mathbb I}_4 \right)\\
& = &
 - {\mathbb I}_4 \partial_0^2
+  e^{-2Ht} {\mathcal A} (x, \partial_x)  {\mathbb I}_4  
 - \left(b-a-1-\frac{c }{2} \right) H e^{-Ht} \gamma ^0 \gamma ^k A_k(x, \partial_x) 
  \\
&  &
-  \left(m {\mathbb I}_4-i\frac{bH}{2} \gamma ^0\right)^2     -i  \left(a+\frac{c}{2} \right)H m \gamma ^0    
+  \left(a-\frac{c}{2} \right) H  {\mathbb I}_4 \partial_0- ( bc+2ab -2ac)\frac{H^2}{4} {\mathbb I}_4\,.  
\end{eqnarray*}
In particular,\\
(i) If $a=c=0=m$   and $b=1$ the second order operator has   the diagonal  imaginary mass $iH/2{\mathbb I}_4$. 
\begin{eqnarray*}
&  &
 \left(i \gamma ^0 \partial_0+ i e^{-Ht} \gamma ^k A_k(x, \partial_x) + i \frac{H}{2}  \gamma ^0  \right) \left(i \gamma ^0 \partial_0+ i e^{-Ht} \gamma ^j A_j(x, \partial_x)-i \frac{H}{2}  \gamma ^0  \right)\\
& = &
  \left(  -  \partial_t^2+ e^{-2Ht}   {\mathcal A}(x, \partial_x)   
 + \frac{  H^2}{4} \right) {\mathbb I}_4   \,.
\end{eqnarray*}
(ii) If $a=1$, $b=3$ and $c=2$, then   
\begin{eqnarray*}
&  &
e^{Ht}\left(i \gamma ^0 \partial_0+ i e^{-Ht} \gamma ^k A_k(x, \partial_x) + i3\frac{H}{2}  \gamma ^0 -m {\mathbb I}_4\right)\\
&  &
\times e^{-Ht}\left(i \gamma ^0 \partial_0+ i e^{-Ht} \gamma ^j A_j(x, \partial_x)-i \frac{H}{2}  \gamma ^0+m {\mathbb I}_4 \right)\\
& = &
    - {\mathbb I}_4  \partial_t^2+ e^{-2Ht} {\mathbb I}_4 {\mathcal A}(x, \partial_x) -\left(m{\mathbb I}_4-\frac{i H}{2}\gamma ^0 \right)^2\,,   
\end{eqnarray*}
where the last  second order scalar operator  represents the generalized  non-covariant Klein-Gordon operators  in de~Sitter spacetime with the complex masses, while 
the first  factor 
is the generalized Dirac operator in de~Sitter spacetime.
\\
(iii) If $a= -1/2$ and $b=3 $, $c= 5$ then    
\begin{eqnarray*}
&  &
e^{ -Ht/2}\left(i \gamma ^0 \partial_0+ i e^{-Ht} \gamma ^k A_k(x, \partial_x) + i3\frac{H}{2}  \gamma ^0 -m {\mathbb I}_4\right)\\
&  &
\times e^{ Ht/2}\left(i \gamma ^0 \partial_0+ i e^{-Ht} \gamma ^j A_j(x, \partial_x)+i2\frac{H}{2}  \gamma ^0+m {\mathbb I}_4 \right)\\
& = & 
 - {\mathbb I}_4 \partial_0^2-3 H  {\mathbb I}_4 \partial_0
+  e^{-2Ht} {\mathcal A} (x, \partial_x) 
    - \left(m {\mathbb I}_4-i\frac{H}{2}\gamma ^0 \right)^2-\frac{9H^2}{4} {\mathbb I}_4\,,
\end{eqnarray*}
where the   second scalar order operators are the generalized covariant Klein-Gordon operators in de~Sitter spacetime  
and  the  diagonal complex valued mass matrix is $\left(m{\mathbb I}_4-\frac{i H}{2}\gamma ^0 \right)^2 +\frac{9H^2}{4}{\mathbb I}_4 $. 
 \end{proposition}
\medskip

\noindent
{\bf Proof.} We start with the case of diagonal   imaginary mass.
First we verify
\begin{eqnarray*}
&  &
-\left(i \gamma ^0 \partial_0+ i e^{-Ht} \gamma ^k A_k(x, \partial_x) + ib\frac{H}{2}  \gamma ^0 \right)\left(i \gamma ^0 \partial_0+ i e^{-Ht} \gamma ^j A_j(x, \partial_x)-ib\frac{H}{2}  \gamma ^0 \right)\\
& = &
  {\mathbb I}_4 \partial_0^2
-  e^{-2Ht} {\mathcal A} (x, \partial_x)    
+ (b-1) H e^{-Ht} \gamma ^0 \gamma ^b A_b(x, \partial_x) 
-  \frac{b^2H^2}{4}{\mathbb I}_4    \,.
\end{eqnarray*}
Indeed, straightforward calculations   lead to the claimed statement: 
\begin{eqnarray*}
  &  &
-\left(i \gamma ^0 \partial_0+ i e^{-Ht} \gamma ^k A_k(x, \partial_x)+ ib\frac{H}{2}  \gamma ^0 \right)
\left(i \gamma ^0 \partial_0+ i e^{-Ht} \gamma ^j A_j(x, \partial_x)-ib\frac{H}{2}  \gamma ^0 \right) \\
 & = &
 \left( \gamma ^0 \partial_0+   e^{-Ht} \gamma ^k A_k(x, \partial_x) \right)\left(  \gamma ^0 \partial_0+   e^{-Ht} \gamma ^j A_j(x, \partial_x) \right) \\
 &  &
 -  \left( \gamma ^0 \partial_0+   e^{-Ht} \gamma ^k A_k(x, \partial_x) \right)  b\frac{H}{2}\gamma ^0 
 +   b\frac{H}{2} \gamma ^0(  \gamma ^0 \partial_0+   e^{-Ht} \gamma ^j A_j(x, \partial_x))
-  \left(b\frac{H}{2} \gamma ^0\right)  ^2  \\
  & = &
 {\mathbb I}_4 \partial_0^2+   \gamma ^0 \partial_0 e^{-Ht} \gamma ^k A_k(x, \partial_x)  +     e^{-Ht} \gamma ^k A_k(x, \partial_x)  \gamma ^0 \partial_0
+  e^{-Ht} \gamma ^k A_k(x, \partial_x) e^{-Ht} \gamma ^j A_j(x, \partial_x) \\
   &  &
  -    \gamma ^0 \partial_0 b\frac{H}{2}\gamma ^0-  e^{-Ht} \gamma ^k A_k(x, \partial_x)   b \frac{H}{2}\gamma ^0 
 +  b \frac{H}{2} {\mathbb I}_4 \partial_0+  b\frac{H}{2} \gamma ^0 e^{-Ht} \gamma ^k A_k(x, \partial_x) 
-   \frac{b^2H^2}{2} {\mathbb I}_4  \\ 
  & = &
 {\mathbb I}_4 \partial_0^2+   \partial_0 e^{-Ht} \gamma ^0 \gamma ^k A_k(x, \partial_x)  +     e^{-Ht}  \gamma ^k \gamma ^0  A_k(x, \partial_x)\partial_0
+  e^{-2Ht} \gamma ^k A_k(x, \partial_x)  \gamma ^j A_j(x, \partial_x)  \\
  &  &
 -    b \frac{H}{2}{\mathbb I}_4 \partial_0 -  e^{-Ht}   b\frac{H}{2}  \gamma ^k  \gamma ^0  A_k(x, \partial_x) 
 +   b\frac{H}{2} {\mathbb I}_4 \partial_0+  b\frac{H}{2} e^{-Ht} \gamma ^0 \gamma ^k A_k(x, \partial_x) 
-   \frac{b^2H^2}{4} {\mathbb I}_4\\
  & = &
  {\mathbb I}_4 \partial_0^2
-  e^{-2Ht} {\mathcal A} (x, \partial_x)    
 + (b-1) H e^{-Ht} \gamma ^0 \gamma ^k A_k(x, \partial_x) 
-   \frac{b^2H^2}{4}{\mathbb I}_4  \,.
\end{eqnarray*}
If $b=1$ the last equation implies (i).  Since $m {\mathbb I}_4$ commutes with any operator, we derive 
\begin{eqnarray*}
\hspace{-0.3cm} &  &
\left(i \gamma ^0 \partial_0+ i e^{-Ht} \gamma ^k A_k(x, \partial_x) + ib\frac{H}{2}  \gamma ^0 -m {\mathbb I}_4\right)\left(i \gamma ^0 \partial_0+ i e^{-Ht} \gamma ^j A_j(x, \partial_x)-ib\frac{H}{2}  \gamma ^0+m {\mathbb I}_4 \right)\\
\hspace{-0.3cm} &  &=
 - {\mathbb I}_4 \partial_0^2
+  e^{-2Ht} {\mathcal A} (x, \partial_x)    
 - (b-1) H e^{-Ht} \gamma ^0 \gamma ^k A_k(x, \partial_x) 
-  \left(m {\mathbb I}_4-i\frac{bH}{2} \gamma ^0\right)^2   \,.
\end{eqnarray*}
Then, for every $c \in {\mathbb C}$ we have 
\begin{eqnarray*}
&  &
\left(i \gamma ^0 \partial_0+ i e^{-Ht} \gamma ^k A_k(x, \partial_x) + ib\frac{H}{2}  \gamma ^0 -m {\mathbb I}_4\right)\\
&  &
\times \left(i \gamma ^0 \partial_0+ i e^{-Ht} \gamma ^k A_k(x, \partial_x)-i(b-c)\frac{H}{2}  \gamma ^0+m {\mathbb I}_4 \right)\\
& = &
 - {\mathbb I}_4 \partial_0^2
+  e^{-2Ht} {\mathcal A} (x, \partial_x)    
 - (b-1) H e^{-Ht} \gamma ^0 \gamma ^k A_k(x, \partial_x) 
- \left(m {\mathbb I}_4-i\frac{bH}{2} \gamma ^0\right)^2  \\
&  &
+ic\frac{H}{2}\left(i \gamma ^0 \partial_0+ i e^{-Ht} \gamma ^k A_k(x, \partial_x) + ib\frac{H}{2}  \gamma ^0 -m {\mathbb I}_4\right)\gamma ^0\\
& = &
 - {\mathbb I}_4 \partial_0^2
+  e^{-2Ht} {\mathcal A} (x, \partial_x)    
 - \left(b-1-\frac{c }{2} \right) H e^{-Ht} \gamma ^0 \gamma ^k A_k(x, \partial_x) 
-   \left(m {\mathbb I}_4-i\frac{bH}{2} \gamma ^0\right)^2   \\
&  &
- c\frac{H}{2}  {\mathbb I}_4  \partial_0    - bc\frac{H^2}{4} {\mathbb I}_4 -i c\frac{H}{2}m \gamma ^0\,.
\end{eqnarray*}
Furthermore, 
\begin{eqnarray*}
&  &
e^{aHt}\left(i \gamma ^0 \partial_0+ i e^{-Ht} \gamma ^k A_k(x, \partial_x) + ib\frac{H}{2}  \gamma ^0 -m {\mathbb I}_4\right) \\
& = & 
\left(i \gamma ^0 \partial_0+ i e^{-Ht} \gamma ^k A_k(x, \partial_x) + ib\frac{H}{2}  \gamma ^0 -m {\mathbb I}_4\right) e^{aHt}
-i \gamma ^0aHe^{aHt}
\end{eqnarray*}
implies
\begin{eqnarray*}
&  &
e^{aHt}\left(i \gamma ^0 \partial_0+ i e^{-Ht} \gamma ^k A_k(x, \partial_x) + ib\frac{H}{2}  \gamma ^0 -m {\mathbb I}_4\right)\\
&  &
\times e^{-aHt}\left(i \gamma ^0 \partial_0+ i e^{-Ht} \gamma ^j A_j(x, \partial_x)-i(b-c)\frac{H}{2}  \gamma ^0+m {\mathbb I}_4 \right)\\
& = &
\Bigg[\left(i \gamma ^0 \partial_0+ i e^{-Ht} \gamma ^k A_k(x, \partial_x) + ib\frac{H}{2}  \gamma ^0 -m {\mathbb I}_4\right)  
-i \gamma ^0aH  \Bigg]\\
& &
\times \left(i \gamma ^0 \partial_0+ i e^{-Ht} \gamma ^j A_j(x, \partial_x)-i(b-c)\frac{H}{2}  \gamma ^0+m {\mathbb I}_4 \right)\\
& = &
\left(i \gamma ^0 \partial_0+ i e^{-Ht} \gamma ^k A_k(x, \partial_x) + ib\frac{H}{2}  \gamma ^0 -m {\mathbb I}_4\right)  
\\
& &
\times \left(i \gamma ^0 \partial_0+ i e^{-Ht} \gamma ^j A_j(x, \partial_x)-i(b-c)\frac{H}{2}  \gamma ^0+m {\mathbb I}_4 \right)\\
&  &
-i \gamma ^0aH  \left(i \gamma ^0 \partial_0+ i e^{-Ht} \gamma ^k A_k(x, \partial_x)-i(b-c)\frac{H}{2}  \gamma ^0+m {\mathbb I}_4 \right).
\end{eqnarray*}
The last expression can be written as follows
\begin{eqnarray*}
&   &
 - {\mathbb I}_4 \partial_0^2
+  e^{-2Ht} {\mathcal A} (x, \partial_x)    
 - \left(b-1-\frac{c }{2} \right) H e^{-Ht} \gamma ^0 \gamma ^k A_k(x, \partial_x) 
-  \left(m {\mathbb I}_4-i\frac{bH}{2} \gamma ^0\right)^2    \\
&  &
- c\frac{H}{2}  {\mathbb I}_4  \partial_0    - bc\frac{H^2}{4} {\mathbb I}_4 -i c\frac{H}{2}m \gamma ^0\\
&  &
-i \gamma ^0aH  \left(i \gamma ^0 \partial_0+ i e^{-Ht} \gamma ^a A_a(x, \partial_x)-i(b-c)\frac{H}{2}  \gamma ^0+m {\mathbb I}_4 \right)\\
& = &
 - {\mathbb I}_4 \partial_0^2
+  e^{-2Ht} {\mathcal A} (x, \partial_x)    
 - \left(b-a-1-\frac{c }{2} \right) H e^{-Ht} \gamma ^0 \gamma ^k A_k(x, \partial_x) 
  \\
&  &
-  \left(m {\mathbb I}_4-i\frac{bH}{2} \gamma ^0\right)^2     -i  \left(a+\frac{c}{2} \right)H m \gamma ^0     
+  \left(a-\frac{c}{2} \right) H  {\mathbb I}_4 \partial_0- ( bc+2ab -2ac)\frac{H^2}{4} {\mathbb I}_4  \,.
\end{eqnarray*}
The proposition is proved. 
\qed

\begin{proposition}
The following identity holds
\begin{eqnarray*}
&  &
 \left(i \gamma ^0 \partial_0+ i e^{-Ht} \gamma ^k A_k(x, \partial_x) + i \frac{3H}{2}  \gamma ^0 -m {\mathbb I}_4\right)\\
&  &
\times  \left(i \gamma ^0 \partial_0+ i e^{-Ht} \gamma ^j A_j(x, \partial_x) + i \frac{3H}{2}  \gamma ^0 +m {\mathbb I}_4\right)\\
& = &
   -    {\mathbb I}_4\partial_0^2- 3 H   {\mathbb I}_4  \partial_0
+  e^{-2Ht}  {\mathbb I}_4{\mathcal A} (x, \partial_x)-He^{-Ht} \gamma ^k \gamma ^0A_k(x, \partial_x) 
-  m^2 {\mathbb I}_4   -\frac{9H^2}{4}  {\mathbb I}_4  \,. 
\end{eqnarray*}
\end{proposition}
\medskip

\ndt
{\bf Proof.} It suffice to consider the case of $m=0$. According to Proposition~\ref{P3.1} case (i):
\begin{eqnarray*}
&  &
 \left(i \gamma ^0 \partial_0+ i e^{-Ht} \gamma ^a A_a(x, \partial_x) + i \frac{3H}{2}  \gamma ^0  \right) \left(i \gamma ^0 \partial_0+ i e^{-Ht} \gamma ^a A_a(x, \partial_x) + i \frac{3H}{2}  \gamma ^0  \right)\\
& = &  
 \left(  -  \partial_t^2+ e^{-2Ht}   {\mathcal A}(x, \partial_x) 
 + \frac{  H^2}{4} \right) {\mathbb I}_4  
+ \left(i \gamma ^0 \partial_0+ i e^{-Ht} \gamma ^a A_a(x, \partial_x) + i \frac{ H}{2}  \gamma ^0 \right)  i2 H \gamma ^0 \\
&  &
+  i H \gamma ^0 \left(i \gamma ^0 \partial_0+ i e^{-Ht} \gamma ^a A_a(x, \partial_x) - i \frac{ H}{2}  \gamma ^0 \right)- 2H^2 {\mathbb I}_4  \\ 
& = &  
 \left(  -  \partial_t^2+ e^{-2Ht}   {\mathcal A}(x, \partial_x) 
 + \frac{  H^2}{4} \right) {\mathbb I}_4  \\  
&  &
+ \left(i \gamma ^0 \partial_0 + i \frac{ H}{2}  \gamma ^0 \right)  i2 H \gamma ^0 
+ \left(  i e^{-Ht} \gamma ^a A_a(x, \partial_x)  \right)  i2 H \gamma ^0\\
&  &
+  i H \gamma ^0 \left(i \gamma ^0 \partial_0  - i \frac{ H}{2}  \gamma ^0 \right)
+  i H \gamma ^0 \left(  i e^{-Ht} \gamma ^a A_a(x, \partial_x)   \right)- 2H^2  {\mathbb I}_4  \\ 
& = &  
 \left(  -  \partial_t^2+ e^{-2Ht}   {\mathcal A}(x, \partial_x) 
 + \frac{  H^2}{4} \right) {\mathbb I}_4  -   He^{-Ht}     \gamma ^a A_a(x, \partial_x)     \gamma ^0-3H {\mathbb I}_4\partial_0 
-     \frac{5}{2}   H^2  {\mathbb I}_4   \,.
\end{eqnarray*}
Proposition is proved.
\qed

\begin{corollary}
If we denote the generalized spinorial covariant D'Alembert operator 
\[
\Box_g=   {\mathbb I}_4\partial_0^2+ 3 H {\mathbb I}_4   \partial_0
-e^{-2Ht} {\mathbb I}_4{\mathcal A} (x, \partial_x)+He^{-Ht} \gamma ^k \gamma ^0A_k(x, \partial_x) 
  -\frac{3H^2}{4}   {\mathbb I}_4  
\] 
and the curvature of the de Sitter spacetime $R=-12H^2$, then
\begin{eqnarray}
&  &
 \left(i \gamma ^0 \partial_0+ i e^{-Ht} \gamma ^k A_k(x, \partial_x) + i \frac{3H}{2}  \gamma ^0  \right) 
\left(i \gamma ^0 \partial_0+ i e^{-Ht} \gamma ^j A_j(x, \partial_x) + i \frac{3H}{2}  \gamma ^0  \right) \nonumber \\
\label{2.18}
& = &
-\Box_g  
-\frac{R}{4} {\mathbb I}_4\,.   
\end{eqnarray} 
\end{corollary}
The   matrices $\gamma ^k \gamma ^0$ in the expression of  $\Box_g $ are not  diagonal. The    formula (\ref{2.18}) is an extension of the relation  (\ref{0.3}) to the generalized Dirac operators in the de~Sitter spacetime.  
\medskip

\ndt
{\bf Example 1.} For the Dirac operator  in the Cartesian coordinates  with the operators   
\begin{eqnarray*}
&  &
 A_k(x, \partial_x) =  \frac{\partial}{\partial x_k},\quad  k=1,2,3, \quad {\mathcal A}(x, \partial_x)  =\Delta 
\end{eqnarray*}
in the spatial part the factorization (\ref{AA}) holds.

\ndt
{\bf Example 2.} Consider the operators
\begin{eqnarray*}
 A_1(x, \partial_x) 
& = &
a (x ,y) \frac{\partial}{\partial x },\quad A_2(x, \partial_x) =b(x,y) \frac{\partial}{\partial y},\quad  A_3(x, \partial_x) =c(z) \frac{\partial}{\partial z}, \\
 {\mathcal A}(x, \partial_x)
&  = &  
\left(a^2(x,y)  \frac{\partial ^2}{\partial x^2} (x,y,z)+b^2(x,y)  \frac{\partial ^2}{\partial y^2}+c^2(z) \frac{\partial ^2}{\partial z^2}\right)   \\
&  &
 +  
  \frac{1}{2}  \left(     \frac{\partial a^2(x,y)}{\partial x }  
  \frac{\partial}{\partial x } 
+  \frac{\partial b^2(x,y)}{\partial y }  
  \frac{\partial}{\partial y}
+  \frac{\partial c^2(z)}{\partial z }  
  \frac{\partial}{\partial z}\right) \,, \\
{\mathcal C}(x, \partial_x) 
& = &
- \left( a_{y}(x,y)b(x,y)\frac{\partial}{\partial x } -  a(x,y) b_{x}(x,y)\frac{\partial}{\partial y } \right) 
\end{eqnarray*} 
with the real-valued smooth coefficients $a(x,y) $, $b(x,y)$, $c(z)$. They  form the generalized Dirac operator.  Indeed, using the properties of the Dirac matrices we verify the condition (\ref{CAA}) as follows
\[
\left( \gamma ^1 a (x ,y) \frac{\partial}{\partial x }+ \gamma ^2 b(x,y) \frac{\partial}{\partial y}+
\gamma ^3 c(z) \frac{\partial}{\partial z}  \right)^2 \\
   =  
 - {\mathcal A}(x, \partial_x){\mathbb I}_4+ {\mathcal C}(x, \partial_x) \gamma ^1 \gamma ^2\,.
\]
If $a=a(x ) $ and $b=b(  y)$, then ${\mathcal C}(x, \partial_x) $ vanishes. If the functions $a(x,y) $, $b(x,y)$, $c(z)$, have zeros, then the Klein-Gordon operator of (\ref{6.17}) is weakly hyperbolic and,  in general, for the well-posedness of the Cauchy problem the so-called Levi conditions are necessary. For  this example they are fulfilled. The  constructions of the parametrix and the fundamental solutions (propagators) for such operators are given in \cite[Ch.4]{Yagbook}.   

\medskip

\ndt
{\bf Example 3.}  {\it The factorization of the Klein-Gordon operator in the de~Sitter spacetime in the cylindrical  coordinates.} 
In the cylindrical coordinates 
\[
x=\rho  \cos (\varphi ),\quad y=\rho  \sin (\varphi ), \quad z=z
\]
one can choose 
\[
A_1(x, \partial ) 
   = 
\cos ( \phi )    \partial_ \rho    
-\frac{ \sin (\phi  )    }{\rho } \partial_ \phi      \, , \,\,
A_2(x, \partial ) 
   =  
\sin (\phi )    \partial_\rho   
+\frac{ \cos (\phi )    }{\rho } \partial_  \phi  \,, \,\, 
A_3(x, \partial ) 
   =    
  \partial_z      \,.
\]
Here  $x_1=r$, $x_2=\phi   $, and $x_3=z$. It is easy to verify that with 
\begin{eqnarray*} 
   {\mathcal A} (x, \partial_x) 
& =  &  
\partial_\rho ^2   + \frac{1}{\rho } \partial_\rho +\frac{1}{r^2}  \partial_\phi  ^2    +  \partial_z  ^2   
\end{eqnarray*} 
 the condition (\ref{AA}) of Proposition~\ref{P3.1} is fulfilled. The Dirac operator is
\begin{eqnarray*}
&  &
 i \gamma ^0 \partial_t 
 +ie^{-H t} \gamma ^1
A_1(x, \partial )
+i e^{-H t} \gamma ^2
A_2(x, \partial ) 
 +ie^{-H t} \gamma ^3
A_3(x, \partial )
+i \frac{3}{2} H  \gamma ^0-m {\mathbb I}_4\,.
\end{eqnarray*}

\ndt
{\bf Example 4. } {\it The factorization of the Klein-Gordon operator in the de~Sitter spacetime in the spherical coordinates.} For the Laplace  operator in the spherical coordinates in ${\mathbb R}^3$ 
\begin{eqnarray*} 
x(r,\theta ,\phi )
  :=  
r \cos (\phi ) \sin (\theta ),\quad
y(r,\theta ,\phi )
  :=  
r \sin (\phi ) \sin (\theta ),\quad
z(r,\theta ,\phi )
  :=  
r \cos (\theta )
\end{eqnarray*} 
one can choose 
\begin{eqnarray*}
A_1(x, \partial ) 
&  = &
\cos ( \phi ) \sin (\theta)  \partial_ r   
+\frac{ \cos (\phi  ) \cos (\theta)  }{r} \partial_ \theta   -\frac{\sin (\phi ) }{r \sin (\theta )} \partial_ \phi  \, ,
 \\ 
A_2(x, \partial ) 
&  = &
\sin (\phi ) \sin (\theta )  \partial_r  
+\frac{ \sin (\phi ) \cos (\theta )  }{r} \partial_  \theta  
+\frac{\cos (\phi ) }{r \sin (\theta )} \partial_ \phi\,,\\ 
A_3(x, \partial ) 
&  = &  
\cos (\theta )  \partial_r   -\frac{\sin (\theta ) }{r}  \partial_\theta   \,.
\end{eqnarray*}
Here  $x_1=r$, $x_2=\theta $, and $x_3=\phi $. It is easy to verify that with 
\begin{eqnarray*}
   {\mathcal A} (x, \partial_x) 
& =  &  
\partial_r^2   + \frac{2}{r} \partial_r+\frac{1}{r^2} \left( \partial_\theta ^2 +\cot (\theta ) \partial_\theta   +\csc ^2(\theta ) \partial_\phi  ^2   
\right)
\end{eqnarray*} 
 the condition (\ref{AA}) of Proposition~\ref{P3.1} is fulfilled. The Dirac operator is
\begin{eqnarray*}
&  &
 i \gamma ^0 \partial_t 
 +ie^{-H t} \gamma ^1
A_1(x, \partial )
+i e^{-H t} \gamma ^2
A_2(x, \partial ) 
 +ie^{-H t} \gamma ^3
A_3(x, \partial )
+i \frac{3}{2} H  \gamma ^0-m {\mathbb I}_4\,.
\end{eqnarray*} 
We can write the Dirac equation
\begin{eqnarray*}
&  &
 \left(i {\gamma }^0  \partial_0+i  e^{-Ht} {\gamma }^a\partial_a  + i       \frac{3}{2}  H  {\gamma }^0   -m{\mathbb I}_4\right)\Psi (x,t)=F(x,t) 
\end{eqnarray*}
as follows
\begin{eqnarray*}
&  &
 \left(i {\gamma }^0  \partial_0+i   e^{-Ht} \left( {\gamma }^r_c \partial_r  +{\gamma }_c^\theta  \partial_\theta+ {\gamma }_c^\phi  \partial_\phi  \right)  + i       \frac{3}{2}  H  {\gamma }^0   -m{\mathbb I}_4\right)\Psi (x,t)=F(x,t) \,,
\end{eqnarray*}
where in this Cartesian tetrad gauge the gamma matrices will be given by (see, e.g., \cite{Schluter}) 
\begin{eqnarray}
\label{3.14}
{\gamma }^r_c
& = &
\gamma ^1 \cos (\phi ) \sin (\theta )
+\gamma ^2 \sin (\theta ) \sin (\phi ) 
+ \gamma ^3  \cos (\theta )\,,\\
\label{3.15}
{\gamma }^\phi _c 
& = &
 -\gamma ^1\frac{\sin (\phi ) }{r \sin (\theta )} 
+ \gamma ^2 \frac{\cos (\phi ) }{r \sin (\theta )} 
=  \frac{ 1}{r \sin (\theta  )}\left( -\gamma ^1\sin (\phi  ) 
+ \gamma ^2 \cos (\phi  ) \right)  \,,\\
\label{3.16}
{\gamma }^\theta _c 
& = &
\frac{ 1 }{r}\left(\gamma ^1\cos (\theta ) \cos (\phi ) 
+ \gamma ^2\sin (\phi) \cos (\theta  ) 
-\gamma ^3 \sin (\theta  )    \right)\,.
\end{eqnarray}
We have used the subscript $c$ for Cartesian. 
We can also write
\begin{eqnarray*}
&  &
 \left(i {\gamma }^0  \partial_0+i   e^{-Ht} \left( \tilde{\gamma }_c^r \partial_r 
 +\tilde{\gamma }_c^\phi  \frac{ 1}{r \sin (\theta)} \partial_\phi
+ \tilde{\gamma }_c^\theta  \frac{ 1 }{r} \partial_\theta  \right)  
+ i       \frac{3}{2}  H  {\gamma }^0   -m{\mathbb I}_4\right)\Psi (x,t)=F(x,t) \,,
\end{eqnarray*} 
where  $\tilde{\gamma }^t_c = {\gamma }^0$ and 
\begin{eqnarray*}
\tilde{\gamma }^r_c
& = &
{\gamma }^r_c=\gamma ^1 \cos (\phi ) \sin (\theta )
+\gamma ^2 \sin (\theta ) \sin (\phi ) 
+ \gamma ^3  \cos (\theta ) \,,\\
\tilde{\gamma }^\phi_c 
& = &
 -\gamma ^1\sin (\phi  ) 
+ \gamma ^2 \cos (\phi  )\,, \\
\tilde{\gamma }^\theta_c 
& = &
\gamma ^1\cos (\theta ) \cos (\phi ) 
+ \gamma ^2\sin (\phi) \cos (\theta  ) 
-\gamma ^3 \sin (\theta  )  \,.    
\end{eqnarray*}
and with the Lorenzian metric $\eta  $ in the Minkowski space time, we have
\begin{eqnarray*} 
&  &
\left\{  \tilde{\gamma}^{\mu}_c,  \tilde{\gamma}^{\nu}_c \right\} = 2 \eta ^{\mu \nu} ,\quad \mu ,\nu = t,r,\theta ,\phi,\quad
 \eta ^{r r}=\eta ^{\theta  \theta } =\eta ^{\phi \phi  } =-1,\quad \eta ^{\mu \nu}=0\quad  if \quad \mu \not= \nu\,. 
\end{eqnarray*}
\begin{proposition}
In the spherical coordinates the following factorization of the Klein-Gordon operator in the de~Sitter spacetime holds
\begin{eqnarray*} 
&  &    
\left(  \partial_0^2     
-  e^{- 2Ht}    
\left(   \frac{ \partial^2 }{ \partial r^2}  
+\frac{2}{r } \frac{ \partial }{ \partial r} 
+ \frac{1}{r^2 } \frac{ \partial^2 }{ \partial \theta ^2 }    
+ \frac{\cot (\theta ) }{r^2 } \frac{ \partial }{ \partial \theta  }  
 + \frac{1}{r^2\sin^2 (\theta )} \frac{ \partial^2 }{ \partial \phi ^2  }  \right)\right){\mathbb I}_4 \\   
&  &
+ \left( m{\mathbb I}_4  - \frac{1}{2}iH \gamma ^0\right)^2 \\
& = &
-e^{Ht}  \left(i {\gamma }^0  \partial_0+i   e^{-Ht} \left( \tilde{\gamma }_c^r \partial_r 
 +\tilde{\gamma }_c^\phi  \frac{ 1}{r \sin (\theta  )} \partial_\phi 
+ \tilde{\gamma }_c^\theta  \frac{ 1 }{r} \partial_\theta   \right)  
+ i       \frac{3}{2}  H  {\gamma }^0   -m{\mathbb I}_4\right) \\
&  &
\times e^{-Ht}\left( i\gamma ^0 \partial_0+  ie^{-Ht} \left(\tilde{\gamma }_c^r \partial_r 
 +\tilde{\gamma }_c^\phi  \frac{ 1}{r \sin (\theta  )} \partial_\phi 
+ \tilde{\gamma }_c^\theta  \frac{ 1 }{r} \partial_\theta   \right)- i\frac{H}{2}\gamma ^0+m{\mathbb I}_4 \right) \,.
\end{eqnarray*}
\end{proposition}
In particular, for the Dirac equation with the source term $ F$ we have 
\begin{eqnarray*}
&  &
 i \partial_0 \Psi +i   e^{-Ht} {\gamma }^0
\left( {\gamma }^r \partial_r + {\gamma }^\phi  \partial_\phi +{\gamma }^\theta  \partial_\theta  \right) + i       \frac{3}{2}  H {\mathbb I}_4   -m{\gamma }^0 \Psi  ={\gamma }^0 F\,,
\end{eqnarray*}
where in this Cartesian tetrad gauge the gamma matrices are given by (\ref{3.14}),(\ref{3.15}),(\ref{3.16}).

\medskip

\ndt
{\bf Example 5.} Let the functions $a (t,x,y,z)$, $b (t,x,y,z)$, $c (t,x,y,z)$, $d (t,x,y,z)$  be such that 
 \begin{eqnarray*} 
&  &
a_{z}(t,x,y,z) =c_{x}(t,x,y,z) ,\quad  
  b_{z}(t,x,y,z)= c_{y}(t,x,y,z)  ,\\
&  &  
a_{t}(t,x,y,z)=d_{x}(t,x,y,z) , \quad
  b_{t}(t,x,y,z)=  d_{y}(t,x,y,z)  ,\quad  
c_{t}(t,x,y,z)=d_{z}(t,x,y,z) \,. 
\end{eqnarray*} 
It is easy to verify  
 in the Cartesian coordinates with 
 \begin{eqnarray*} 
&  &
 A_0(t,x,y,z ,   \partial_t)
   =   \frac{\partial}{\partial t }+ d (t,x,y,z) ,\quad 
A_1(t,x,y,z ,  \partial_x )
   =   \frac{\partial}{\partial x }+ a (t,x,y,z) ,\\
   &  &
A_2(t,x,y,z ,  \partial_y )
   =   \frac{\partial}{\partial y}+ b(t,x,y,z) ,\quad 
A_3(t,x,y,z ,   \partial_z)
   =   \frac{\partial}{\partial z}+ c(t,x,y,z) ,  
\end{eqnarray*}
for  the Dirac operator 
\begin{equation}  
\label{2.24}
  i\gamma ^0 A_0(t,x,y,z ,   \partial_t)+ i\gamma ^1 A_1(t,x,y,z ,  i\partial_x )+ \gamma ^2 A_2(t,x,y,z ,  \partial_y )+i\gamma ^3 A_3(t,x,y,z ,   \partial_z)-m {\mathbb I}_4 , 
 \end{equation} 
the    following identity  
 \begin{eqnarray*} 
&  &
\left( \gamma ^0 A_0(t,x,y,z ,   \partial_t)+ \gamma ^1 A_1(t,x,y,z ,  \partial_x )+ \gamma ^2 A_2(t,x,y,z ,  \partial_y )+\gamma ^3 A_3(t,x,y,z ,   \partial_z) \right)^2\\ 
& = & 
\Bigg\{\partial_t^2 -\Delta -2\left(   a(t,x,y,z) \frac{\partial}{ \partial x} + b(t,x,y,z) \frac{\partial}{ \partial y}+  c(t,x,y,z) \frac{\partial}{ \partial z} \right) +  2d(t,x,y,z) \frac{\partial}{ \partial t}   \\
&  &
    -a(t,x,y,z)^2  -b(t,x,y,z)^2  -c(t,x,y,z)^2  +d(t,x,y,z)^2 
\\
&  &
  -\frac{\partial a (t,x,y,z)}{ \partial x}   - \frac{\partial b (t,x,y,z)}{ \partial y}  
-\frac{\partial c (t,x,y,z)}{ \partial z}    +\frac{\partial d (t,x,y,z)}{ \partial t}   \Bigg\}{\mathbb I}_4\\
&  &
+ \left(   \frac{\partial a (t,x,y,z)}{ \partial y}   - \frac{\partial b (t,x,y,z)}{ \partial x}  \right) \gamma ^1\gamma ^2 \,.
\end{eqnarray*} 
If $  id (t,x,y,z) , ia (t,x,y,z) ,$ $ ib (t,x,y,z) $, $ic (t,x,y,z) $  are real-valued functions, then the operator (\ref{2.24}) is the {\it Dirac operator 
 in the presence of an electromagnetic potential}  $ i (d (t,x,y,z) $, $a (t,x,y,z) $, $ b (t,x,y,z) $, $c (t,x,y,z))$ for the particle with charge $e=1$.   Thus, the square of this operator is a diagonal matrix of differential operators. It allows us to reduce solution of the corresponding equation to the scalar Klein-Gordon equation. This reduction is effective to produce the explicit formulas of solution.  

  In particular, if $i \vec{A} (x,y,z) =i ( a ( x,y,z) ,  b ( x,y,z),c ( x,y,z))$ is a magnetic potential, then the magnetic field $\vec{H}=i\mbox{\rm curl}  \vec{A} (x,y,z) $ is in the direction of $z$ and time independent. The case of $ a ( x,y,z)=  c ( x,y,z)=0$ and $b( x,y,z)=Hx $, where $H$ is constant, while
 \[ 
\left( i\gamma ^0    \partial_t + i\gamma ^1    \partial_x + \gamma ^2 ( i\partial_y  -Hx  )  +i\gamma ^3     \partial_z  \right)^2 
=
-\left(\partial_t^2 -\Delta -2   Hx \frac{\partial}{ \partial y}      
     -H^2 x^2  \right){\mathbb I}_4 
  + H \gamma ^1\gamma ^2 \,,
\] 
is of particular interest and thoroughly studied in the literature, see, e.g., \cite[Sec.~1.6.2]{Akhiezer}. Using the last formula we can write the explicit formulas for the general solution of the Dirac equation in the de~Sitter spacetime
 \[ 
\left( i\gamma ^0    \partial_t + ie^{-Ht}\gamma ^1    \partial_x + e^{-Ht}\gamma ^2 ( i\partial_y  -Hx  )  +ie^{-Ht}\gamma ^3     \partial_z 
+i\frac{3}{2}H\gamma ^0 -m  {\mathbb I}_4\right) \psi 
= 0 \,.
\]
This will be done in the forthcoming paper. For the irrotational vector field  the term with $\gamma ^1\gamma ^2 $ vanishes.  
 \medskip

\section{Proof of main Theorem~\ref{T0.1}}
\label{S3}

\begin{proposition}
The following factorization of the Klein-Gordon operator with the matrix-valued mass holds
\begin{eqnarray*} 
&  &    
(  \partial_0^2     
-  e^{- 2Ht}    \Delta ){\mathbb I}_4   
+ \left( m{\mathbb I}_4  - \frac{1}{2}iH \gamma ^0\right)^2 \\
& = &
-e^{Ht} \left( i\gamma ^0 \partial_0+ i e^{-Ht} \gamma ^a \partial_a+ i\frac{3H}{2}\gamma ^0-m{\mathbb I}_4 \right) 
e^{-Ht}\left( i\gamma ^0 \partial_0+  ie^{-Ht} \gamma ^a \partial_a- i\frac{H}{2}\gamma ^0+m{\mathbb I}_4 \right) \,.
\end{eqnarray*}
\end{proposition}
\medskip

\noindent
{\bf Proof.} Indeed, by  simple calculations we check the following identity  
\begin{eqnarray*}
&  &
\left(i \gamma ^0 \partial_0+ i e^{-Ht} \gamma ^a \partial_a+ i\frac{H}{2}\gamma ^0-m{\mathbb I}_4 \right)\left(i \gamma ^0 \partial_0+ i e^{-Ht} \gamma ^a \partial_a-i\frac{H}{2}\gamma ^0+m{\mathbb I}_4 \right)\\
& = &
\left(- \square + \frac{1}{4}H^2\right){\mathbb I}_4 +iHm\gamma ^0 - m^2{\mathbb I}_4 \\
& = &
- \square {\mathbb I}_4 - \left( m{\mathbb I}_4  - \frac{1}{2}iH \gamma ^0\right)^2 \,,
\end{eqnarray*}
where
$
\square=\partial_0^2- e^{-2Ht}\Delta =\partial_0^2- e^{-2Ht}\left( \partial_1^2+\partial_2^2+\partial_3^2\right)
$.
 \qed 
\medskip

\noindent
{\bf Proof of Theorem~\ref{T0.1}.} Indeed we can write
\begin{eqnarray*} 
&  &
 \left( i\gamma ^0 \partial_0+  ie^{-Ht} \gamma ^a \partial_a+ i\frac{3H}{2}\gamma ^0-m{\mathbb I}_4\right) 
{\mathcal E}^{ret} \left(x, t ; x_{0}, t_{0};m\right)=  \\ 
& = &
- \left( i\gamma ^0 \partial_0+  ie^{-Ht} \gamma ^a \partial_a+ i\frac{3H}{2}\gamma ^0-m{\mathbb I}_4\right) \\
&  &
\times  e^{-Ht}\left( i\gamma ^0 \partial_0+  ie^{-Ht} \gamma ^a \partial_a- i\frac{H}{2}\gamma ^0+m{\mathbb I}_4\right)  {\mathcal E}_{ret,KG}\left(x, t ; x_{0}, t_{0};m\right)[e^{H\cdot } ] =\\ 
& = &
 e^{-Ht} \left( \partial_0^2     
-  e^{- 2Ht}    \Delta    
+ m ^2  -  \frac{1}{4}H^2 +imH\gamma ^0 \right)    {\mathcal E}_{ret,KG}\left(x, t ; x_{0}, t_{0};m\right)[e^{H\cdot } ]\\ 
& = &
  e^{-Ht} \left(  e^{Ht} \delta (x-x_0)\delta (t-t_0)\right){\mathbb I}_4 \\
& = &
\delta (x-x_0)\delta (t-t_0){\mathbb I}_4\,.
\end{eqnarray*}
Theorem is proved. 
\qed

\medskip

\section{Conclusions}

In this paper, we have used the integral transform approach to write various fundamental solutions to the Dirac equation in the de~Sitter space.   This new integral transform   one can regard  as an analytical mechanism that generates a spin-1/2 (massive or massless) field in the curved spacetime from the   massless scalar field in the Minkowski spacetime.  We   also have shown that  this mechanism exists even in the case of the vanishing cosmological constant; it provides spin-1/2 particles with the mass due to massless scalar field. In fact,  this mechanism appeals to some additional ``time'' variable.

\begin{small}

\end{small}
\end{document}